\documentclass[preprint2]{aastex}
\usepackage{graphicx}
\shorttitle{Sizes of Lyman Alpha Emitters at z=3.1}
\shortauthors{Bond et al.}

\begin{document}

\title{SIZES OF LY$\alpha$-EMITTING GALAXIES \\
AND THEIR REST-FRAME ULTRAVIOLET COMPONENTS AT $z=3.1$\footnote{Based on
    observations made with the NASA/ESA Hubble Space Telescope, and
    obtained from the Hubble Legacy Archive, which is a collaboration
    between the Space Telescope Science Institute (STScI/NASA), the
    Space Telescope European Coordinating Facility (ST-ECF/ESA) and
    the Canadian Astronomy Data Centre (CADC/NRC/CSA). }}

\author{Nicholas A. Bond, Eric Gawiser\altaffilmark{2}, Caryl Gronwall, Robin Ciardullo\altaffilmark{3}, Martin Altmann\altaffilmark{4}, and Kevin Schawinski\altaffilmark{5,6}}
\altaffiltext{2}{Physics and Astronomy Department, Rutgers University
    Piscataway, NJ 08854-8019, U.S.A.}
\email{nbond@physics.rutgers.edu}
\altaffiltext{3}{Department of Astronomy and Astrophysics, Pennsylvania State University, University Park, PA 16802, U.S.A.} 
\altaffiltext{4}{University of Heidelberg, Center for Astronomy, M\"onchhofstr. 12-14, D-69120 Heidelberg, Germany }
\altaffiltext{5}{Department of Physics, Yale University, New Haven, CT 06511, U.S.A.}

\altaffiltext{6}{Yale Center for Astronomy and Astrophysics, Yale University, P.O. Box 208121, New Haven, CT 06520, U.S.A}

\begin{abstract}

  We present a rest-frame ultraviolet analysis of $\sim120$ $z\sim3.1$
  Lyman Alpha Emitters (LAEs) in the Extended Chandra Deep Field South
  (ECDF-S).  Using {\it Hubble Space Telescope} ({\it HST}) images
  taken as part of the Galaxy Evolution From Morphology and SEDS
  (GEMS) survey, Great Observatories Origins Deep Survey (GOODS), and
  Hubble Ultradeep Field (HUDF) surveys, we analyze the sizes of LAEs,
  as well as the spatial distribution of their components, which are
  defined as distinct clumps of UV-continuum emission.  We set an
  upper limit of $\sim 1$~kpc ($\sim 0\farcs1$) on the rms offset
  between the centroids of the continuum and Ly$\alpha$ emission.  The star formation rates of LAE components inferred from the rest-frame ultraviolet continuum
 range from $\sim 0.1$~M$_{\sun}$~yr$^{-1}$ to $\sim 5$~M$_{\sun}$~yr$^{-1}$.  A
  subsample of LAEs with coverage in multiple surveys (at different
  imaging depths) suggests that one needs a signal-to-noise ratio,
  S/N~$\gtrsim 30$, in order to make a robust estimate of the
  half-light radius of an LAE system.  The majority of LAEs have
  observed half-light radii $\lesssim 2$~kpc, and LAE components
  typically have observed half-light radii $\lesssim 1.5$~kpc
  ($\lesssim 0\farcs20$).  Although only $\sim 50$\% of the detected
  LAE components are resolved at GOODS depth, the brightest
  ($V\lesssim 26.3$) are all resolved in both GOODS and GEMS.  Since
  we find little evidence for a correlation between the rest-UV sizes
  and magnitudes of LAEs, the majority should be resolved in a deeper
  survey at the $\sim 0\farcs05$~angular resolution of {\it HST}.
  Most of the multi-component LAEs identified in shallow frames become
  connected in deeper images, suggesting that the majority of the
  rest-UV ``clumps'' are individual star-forming regions within a
  single system.

\end{abstract}

\keywords{cosmology: observations --- galaxies: formation -- galaxies: high-redshift -- galaxies: structure}

\vspace{0.4in}

\section{INTRODUCTION}

In the local universe, the majority of galaxies fall on a sequence
that runs from red, quiescent galaxies with a compact spheroidal component
to star-forming, gas-rich disks with approximately exponential
profiles.  Out to intermediate redshifts ($z \sim 1.5$), there is a clear
continuum in morphological properties that is consistent with the
Hubble Sequence that we observe locally \citep{Cons04}.  However,
at higher redshifts, typical galaxies appear clumpy and irregular
\citep[e.g.,][]{Steidel96,Papovich05,Conselice05,Venemans05,Pirzkal07}
and evade clean placement into existing classification schemes.  

The most studied class of galaxy includes objects found by the
Lyman-break technique, wherein high-redshift galaxies are identified
by a flux discontinuity in the continuum caused by absorption of
intervening neutral hydrogen \citep{Steidel96}.  Morphological
analyses of $z > 2.5$ Lyman-break galaxies (LBGs) have revealed that
most of these systems are disturbed and disk-like (i.e., with
exponential light profiles), with only $\sim 30\%$ having light
profiles consistent with galactic spheroids
\citep[e.g.,][]{Ferguson04,Lotz06,Ravindranath06}.  In addition, using
SExtractor, these studies find a mean half-light radius of $\sim
2.27$~kpc at $z=3.1$ and a size evolution that scales approximately as
$H^{-1}(z)$.

Like LBGs, Lyman Alpha Emitters (LAEs) at $z \sim 2 - 4$ are widely
believed to be actively star-forming \citep[e.g.][]{CowieHu}.
However, they are found to have lower stellar and dark matter masses,
higher mass-specific star-formation rates, and lower dust content on
average \citep{Venemans05,Gawiser07}.  The effort to measure the
morphologies of these objects is still in its earliest stages, with
the majority of the existing results being reported in the broadband
rest-frame ultraviolet.  The qualitative rest-UV morphological
properties of LAEs are generally agreed upon, but LAEs remain
difficult to place in existing classification schemes.  At $3 \lesssim
z \lesssim 6$, most are small (with half-light radii $\lesssim
1$~kpc), compact ($C>2.5$), and barely resolved at {\it Hubble Space
  Telescope} ({\it HST}) resolution
\citep{Venemans05,Pirzkal07,Overzier08,Taniguchi09}.  However, many
($\sim 20 - 45$\%) are clumpy or irregular, with components extending
to several kiloparsecs.

The Multiwavelength Survey by Yale-Chile \citep[MUSYC,][]{MUSYC} is a collaborative effort to obtain multiwavelength imaging and spectroscopy of $1.2$~degree$^2$ of sky in four different fields, including the Extended Chandra Deep Field-South (ECDF-S).  As part of this survey, \citet{GronwallLAE} used broadband and 4990~\AA\ narrow-band imaging of the ECDF-S to identify a large, unbiased sample of
 LAEs at $z=3.1$.  The authors found that their LAE sample had an exponential equivalent width distribution, with a scale length of $w_0=76^{+11}_{-8}$, and followed a Schechter function \citep{Schechter76} in emission-line luminosity, with
 $\alpha=-1.49^{+0.45}_{-0.34}$ and log $L^*=42.64^{+0.26}_{-0.15}$.  In addition, they found that the star formation rates (SFRs) estimated from the UV continuum were $\sim 3$ times larger than those estimated from the Ly$\alpha$ line, with UV SFRs ranging from $\sim 1$ to $10$~M$_{\sun}$~yr$^{-1}$.  Subsequent analysis of this sample by \citet{Gawiser07} showed LAEs to be weakly clustered, with
a bias factor ($b \sim 1.7$) consistent with that expected from the progenitors of present-day $L^*$ galaxies.  Moreover, although $\sim 70\%$ of these LAEs are
 too faint to be detected on deep images taken by the {\it Spitzer} Infrared Array Camera, spectral energy distribution fits to the broadband optical and infrared colors of a mean ``stacked'' LAE suggests that they typically have very small
 stellar masses, $\sim 10^9 M_{\sun}$.

This paper complements the \citet{GronwallLAE} and \citet{Gawiser07} studies of the MUSYC $z=3.1$ LAE sample by measuring the rest-UV size and component distributions using high-resolution $V$-band images, taken by the Advanced Camera for Surveys (ACS) and obtained as part of the Galaxy Evolution from Morphology and SEDs survey \citep[GEMS,][]{GEMS}, Great Observatories Origins Deep Survey \citep[GOODS,][]{GOODS}, and Hubble Ultradeep Field survey \citep[HUDF,][]{HUDF}.  In addition to presenting the largest LAE morphological study to date, this paper describes a new pipeline for the study of high-redshift galaxies at low signal-to-noise.  In past work, clumpy LAEs have been given only crude, qualitative descriptions, but even at low redshift, ordinary late-type galaxies can look clumpy in
 the UV.  Thus, there is ambiguity in the treatment of individual components \citep [][and references therein]{BondRev}.  Here, we fit each photometric component separately and give quantitative size measures for both the individual components {\it and\/} the LAE system as a whole.  Furthermore, since some of our LAEs
are covered in multiple surveys, we present an analysis of the depth dependence of the sizes and component distributions.  As discussed in \citet{BondRev}, this
 is a crucial step if we wish to compare morphological measurements between different high-resolution imaging observations of
LAEs.  We will present an analysis of the higher-order morphological properties of LAE components in a subsequent paper (C. Gronwall et al. 2009, in preparation).

In \S~\ref{sec:data} and \ref{sec:method}, we describe the data and
detail the pipeline used in our analysis.  In \S~\ref{sec:results}, we
present the photometric properties, including half-light radii, of
each LAE system and its components.  We also explore how these
properties vary with image depth.  Finally, in
\S~\ref{sec:discussion}, we discuss the implications of our findings
and suggest a direction for future morphology studies of LAEs and
other high-redshift galaxies.  Throughout this paper, we will assume a
concordance cosmology with $H_0=71$~km~s$^{-1}$~Mpc$^{-1}$,
$\Omega_{\rm m}=0.27$, and $\Omega_{\Lambda}=0.73$ \citep{WMAP}.  With these
values, $1\arcsec = 7.75$~physical~kpc at $z=3.1$.

\begin{deluxetable}{lcccc}
\tablecaption{{\it HST} Imaging Survey Properties\label{tab:SurveyProps}}
\tablewidth{0pt}
\tablehead{
\colhead{Survey}
&\colhead{Sky Coverage}
&\colhead{$V$--band Depth\tablenotemark{a}}
&\colhead{$N_{\rm LAE}$\tablenotemark{b}}
&\colhead{Reference}\\
&
\colhead{(arcmin$^2$)}
&\colhead{(AB mags)}
&
&
}
\startdata
GEMS	&800	&28.3\tablenotemark{c}	&97 &1 \\
GOODS	&160	&28.8	&29 &2 \\
HUDF	&11	&30.5	&4 &3 \\
\enddata
\tablenotetext{a}{$5 \,\sigma$ depth for point source detection}
\tablenotetext{b}{Number of LAEs covered in the survey region}
\tablenotetext{c}{The sGOODS data are shallower, $27.9$~AB~mags}

\tablerefs{
(1) Rix et al. 2004; (2) Giavalisco et al. 2004; (3) Beckwith et al. 2006}

\end{deluxetable}

\section{DATA}
\label{sec:data}

Our study uses the statistically complete sample of $z = 3.1$ LAEs
identified by \citet{GronwallLAE} in the Extended Chandra Deep
Field-South; these objects are defined to have monochromatic fluxes,
$F_{4990} > 1.5 \times 10^{-17}$~ergs~cm$^{-2}$~s$^{-1}$, and
observed-frame Ly$\alpha$ equivalent widths, EW$>80$~\AA\null.  As
published, the \citet{GronwallLAE} sample contains $162$ objects. From
this list, we exclude the two X-ray sources removed by the authors,
one duplicate object (LAE $110$, identical to LAE $124$), and five
recently-identified spurious detections (LAEs $33$, $48$, $57$, $104$,
and $139$, Guillermo Blanc, private communication) caused by CCD
cross-talk in the narrow-band image.  Excluding LAEs within $40$
pixels of the edge of an image, a total of $116$ of the remaining
$154$ objects fall in fields observed by {\it HST}; these are listed
in Table~\ref{tab:SurveyProps}.  Below, we summarize the data.

\subsection{GEMS}
\label{subsec:gems}

The GEMS survey consists of a series of $63$ ACS pointings in the
$V_{606}$ and $z_{814}$-bands, which cover the full $\sim
800$~arcmin$^2$ of the ECDF-S\null.  The depth of this survey is
fairly uniform across the field, with $V_{606}$-band point sources
detected with $5 \,\sigma$ confidence to $m_{\rm AB} = 28.3$ in the
main GEMS survey, and to $m_{\rm AB} = 27.9$ in the region covered by
the first epoch of the GOODS survey (hereafter, sGOODS).  The sGOODS
data were reduced with the GEMS pipeline, but include data
incorporated into the deeper GOODS {\it v2.0} images and will
therefore only be used to test the depth dependence of our
morphological diagnostics (see \S~\ref{subsec:depth}).  All images
have been multidrizzled \citep{Multidrizzle} to a pixel scale of
$0\farcs03$ pixel$^{-1}$ and in the GEMS-only tiles, $97/154$ LAEs are
covered by the survey.

\subsection{GOODS}
\label{subsec:goods}

In the Chandra Deep Field-South, the southern half of the GOODS survey
covers $\sim 160$~arcmin$^2$ of sky and includes {\it HST}/ACS
observations in the $B_{435}$, $V_{606}$, $I_{775}$, and $z_{850}$
filters.  The effective exposure time of this survey is variable across
the GOODS area, but for point sources, a typical $V_{606}$-band,
$5 \,\sigma$ detection limit is $m_{\rm AB} = 28.8$.  All images have
been multidrizzled to a pixel scale of $0\farcs03$ pixel$^{-1}$ and of
$154$ LAEs in our original sample, $29$ have $V_{606}$-band coverage
in {\it v2.0} of the GOODS/ACS catalog.  

\subsection{HUDF}
\label{subsec:hudf}

The images of the Hubble Ultra-Deep Field (HUDF) are deeper than those in either GEMS or GOODS, reaching $V$-band $5 \,\sigma$ point source depth of $m_{\rm AB} = 30.5$, but cover only $11$~arcmin$^2$ of sky.  As in GOODS, the HUDF survey includes {\it
  HST}/ACS observations in the $B_{435}$, $V_{606}$, $I_{775}$, and
$z_{850}$ filters, which have been multidrizzled to a plate scale of
$0\farcs03$ pixel$^{-1}$.  Only $3$ of our $154$ objects
fall in this region.

\section{METHODOLOGY}
\label{sec:method}

High-redshift galaxies frequently exhibit ``clumpy'' morphologies; in such systems, high-order morphological fits can be difficult to interpret.  To avoid this problem, each LAE system was first examined with SExtractor \citep{SExtractor}, to identify individual rest-UV components.  The pipeline developed for this work operated in five stages:
\begin{itemize}
\item{Cutout extraction from survey images (\S~\ref{subsec:cutouts})}
\item{Source detection, using SExtractor (\S~\ref{subsec:cutouts})}
\item{Centroid estimation and aperture photometry using {\tt PHOT} (\S~\ref{subsec:aperture})}
\item{Light profile fitting, using {\tt GALFIT} (\S~\ref{subsec:pointsource})}
\item{Identification of point sources (\S~\ref{subsec:pointsource})}
\end{itemize}

\subsection{Cutouts and SExtractor Runs}
\label{subsec:cutouts}
We began by extracting an $80 \times 80$ pixel ($2\farcs4 \times
2\farcs4$) cutout from the {\it HST}/ACS survey image at the position of
each LAE in our sample.  This region, which has a linear scale of
$\sim 19$~kpc at the redshift of the emitter, is large enough to
include the expected uncertainties in the $V$-band centroids (see
\S~\ref{subsec:aperture}).  Since the profile fits described in
\S~\ref{subsec:pointsource} were performed over the entire cutout, our
final sample included only those LAEs with full survey coverage in the
cutout region.

After extracting the cutouts, we identified all sources contained
within them using the SExtractor \citep{SExtractor} object detection
algorithm.  Since LAEs can appear as either point sources or extended
objects at {\it HST\/} resolution, we set our parameters to find all
sources with at least nine pixels above a $1.65 \,\sigma$ detection
threshold.  Although this condition does not allow us to find very
weak compact sources, even when they are apparent to the eye, this
limitation is not serious, since these objects contain no useful
morphological information.  Finally, to identify those objects with
multiple components, we set the SExtractor parameter, {\tt
  DEBLEND\_MINCONT}$=0.06$; this value was chosen to split the
LAE components which appeared by eye to be separate objects.

\begin{figure}[t]
\plotone{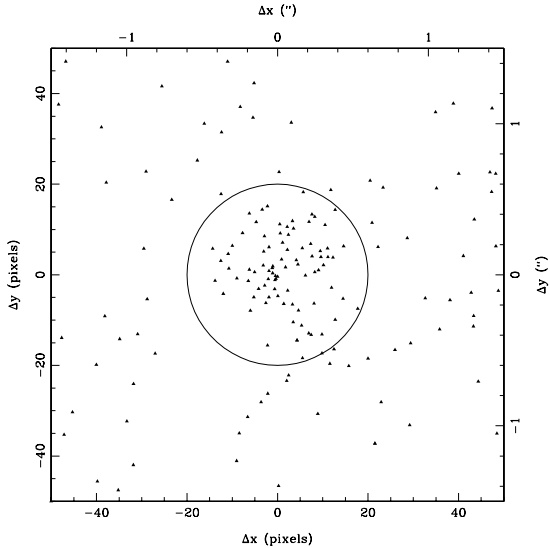}
\caption{Distribution of SExtractor detections in the $V_{606}$-band GEMS
cutouts as a function of distance from the ground-based
Ly$\alpha$ centroid.  Within the $0\farcs6$~($20$-pixel) selection
radius, drawn in black, are objects classified as LAE components.
\label{fig:Exclusion_Radius}}
\end{figure}

Figure~\ref{fig:Exclusion_Radius} plots the distribution of SExtractor
$V_{606}$-band detections in the 97 GEMS cutouts as a function of
angular distance from the ground-based Ly$\alpha$ centroid.  The
detections are highly clustered: $34$ components fall within
$0\farcs25$ of the ground-based position, which is the
approximate positional uncertainty associated with the ground-based
astrometry \citep{MUSYC}.  Moreover, the density of detected
components does not fall to that of the field until $\sim 0\farcs6$,
which we define as our selection radius, $R_{\rm sel}$.  Based on the
density of field sources displayed in
Figure~\ref{fig:Exclusion_Radius}, we estimate that $11$ of the $87$
components detected by SExtractor within our selection radius are
chance coincidences.

After discarding those cutouts with no detections within the selection
radius, we used SExtractor to fit and subtract a uniform sky from each
of the remaining images.  This is a critical step; as a result of
resonant scattering, the diffuse emission from Ly$\alpha$ can extend
many half-light radii beyond the main body of a galaxy
\citep{Ostlin08}.  By using a field cutout size of $2\farcs4$, we
minimize the risk that our estimate of the sky will be affected by
diffuse emission that may occur within our $0\farcs6$ selection
radius.  Similarly, by adopting a uniform sky background, we avoid the
risk of confusing Ly$\alpha$ emission with background fluctuations.

\subsection{Centroid Estimation and Aperture Photometry}
\label{subsec:aperture}
The above procedure is useful for isolating individual components within the LAE cutouts, but we also wish to measure the photometric properties of the composite system; that is, of all light within the selection radius of an LAE.  To estimate the rest-UV centroid of the LAE system, we again run SExtractor on each of the cutouts, now requiring a detection to have only five pixels above the $1.65 \, \sigma$ threshold.  We then measure the centroid to be the flux-weighted mean position of the detections within the selection radius.  The smaller five-pixel detection threshold will find more dim components that, although too dim for a reliable half-light radius determination, could allow for a more accurate determination of the LAE centroid.

We then use the IRAF routine {\tt PHOT}, summing the counts within a
series of apertures, each centered on the measured light centroid and
ranging from $0\farcs015$--$0\farcs6$ in radius.  Assuming that all of
the flux from the LAE system is contained within a $0\farcs6$
aperture, the half-light radius, $r_e^{\rm PHOT}$, is found by
interpolating the curve of growth at one-half of this total flux.  We use
a $0\farcs6$ maximum aperture because it corresponds to the selection
radius (larger maximum apertures yield half-light radii that differ by
no more than $10$\%).

\subsection{Morphology Fits and Point Source Identification}
\label{subsec:pointsource}

We measured the morphological properties (presented in detail in Paper
2) of our LAE sample using {\tt GALFIT} \citep{GALFIT}, a software
package that convolves a model light profile with the point spread
function (PSF) and minimizes $\chi^2$ over a chosen set of model
parameters.  {\tt GALFIT} is fast and capable of simultaneously
fitting multiple sources in a given image, making it an efficient
option for analyzing large samples of multi-component objects.

For the GEMS and GOODS data, we defined the PSFs using a sample of bright stars located throughout the ECDF-S\null.  Only those stars with centroids lying near the center of a pixel and with peak fluxes well below saturation were used in this defintion.  In the case of the extremely small field of the HUDF, only a single star was used for the PSF.  However, since all three LAEs located in the HUDF are well resolved, this limitation is not important for our study.  We then simultaneously fit S\'{e}rsic profiles \citep{Sersic} to all detections within each cutout using elliptical model isophotes.  Unless otherwise specified, we fit to the entire cutout, but only report the properties of a component if its center falls within the LAE selection circle.  No bad pixel masks were used, and each fit was inspected by eye.

The majority of LAEs have half-light radii $<1$~kpc in $V_{606}$
\citep{Venemans05,Pirzkal07,Overzier08}, so many of the objects in our
sample may be unresolved at the $0\farcs06$ ($\sim 0.5$~kpc at
$z=3.1$) resolution typical of {\it HST}.  
To determine whether an object is resolved, we compare the 
$\chi^2$ value of its S\'{e}rsic fit to that of a fit to the PSF alone;
in other words, we require
\begin{equation}
F \equiv \frac{(\chi_{\rm psf}^2-\chi_{\rm sersic}^2)}{\chi_{\rm sersic}^2}>F_{\rm crit}.  
\label{eq:deltachi}
\end{equation}
When data are uncorrelated and have only gaussian random errors, $F_{\rm crit}$ is determined from the F-distribution.  Unfortunately, for point sources, the $\chi^2$ surfaces of the S\'{e}rsic profile are not well behaved, and {\tt GALFIT} (which employs the Levenberg-Marquardt algorithm, see \citet{NumericalRecipes}) does not always converge to the absolute minimum in $\chi^2$.  Consequently, to perform this test, we computed $F_{\rm crit}$ empirically using known stars.

\begin{figure}[t]
\plotone{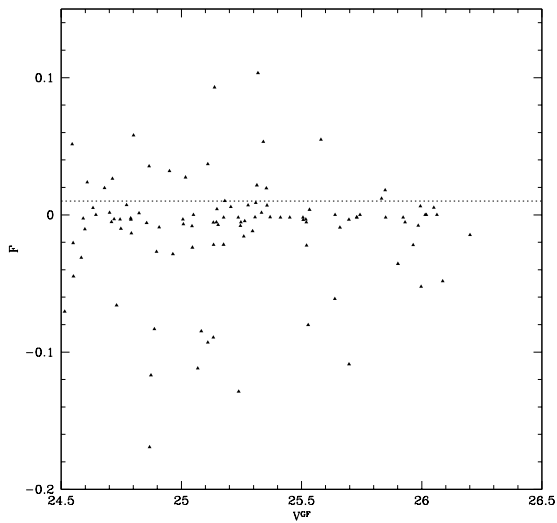}
\caption{Relative improvement of a S\'{e}rsic fit vs.\ a PSF fit 
(represented by $F$-values) for $115$ known point sources (stars), 
plotted as a function of the V-band magnitude output by {\tt GALFIT}.  Here, $\sim 85$\%~of true point
sources have $F<0.01\equiv F_{\rm crit}$ (indicated by the dashed line).
\label{fig:Fcrit}}
\end{figure}

A sample of high-confidence point sources was obtained from
\citet{MartinStars}, who used broadband spectral energy distribution
fits to distinguish stars from galaxies in the ECDF-S\null.  From this
sample, we selected $912$ stars that fall within the GEMS region and
had temperatures, $T<4500$~K, a regime in which photometric confusion
with galaxies is minimal.  After running SExtractor on the GEMS cutout
of each star, we further restricted our sample to 115 objects that
were isolated (i.e., the only object in the cutout), well-centered
(within $0\farcs45$ of the cutout center), unambiguously stellar
(SExtractor stellarity $>0.9$), and faint ($V_{606}^{\rm SE}>24.6$).  This
ensured that we had a clean sample of stars with photometric
uncertainties dominated by sky noise.  Finally, we fit our stellar
sample to both the point spread function and a S\'{e}rsic profile and
plot the resulting $F$ values against V-band magnitude
(Figure~\ref{fig:Fcrit}).  From this plot, we infer that $85$\%~of
true point sources will have $F<0.01\equiv F_{\rm crit}$ (indicated by
the dashed line); we use this threshold to identify LAE components
that are consistent with point sources.

\subsection{Objects with Coverage in Multiple Surveys}
\label{subsec:depth}

Many of the standard measures of a galaxy's morphology exhibit a
systematic offset from their intrinsic values if measured on
low signal-to-noise (S/N) images.  For example, \citet{Ravindranath06}
have fit S\'{e}rsic profiles to a series of models images
with a range of S/N.  At low signal levels, they see a
systematic offset between the input and output S\'{e}rsic index, $n$, where
it is overestimated for model disks and underestimated for model
spheroids.   

Since there are regions of overlap in the sky coverage of the HUDF, GOODS, and GEMS surveys, we can estimate this dependence using a subsample of LAEs in the field.  Specifically, since the HUDF is a subregion of the GOODS survey, all three of the LAEs in that field also have GOODS and sGOODS data.  Similarly, $22/29$ LAEs in GOODS are also present in sGOODS, and there is a small region of overlap between GOODS and GEMS which contains nine LAEs.  We note that there is a systematic offset between the world coordinate systems (WCS) of the GOODS and sGOODS images in the northern part of the Chandra Deep Field-South.  A comparison of the positions of bright sources in each survey shows that the coordinates from GOODS must be shifted by $-7$ pixels in $x$ and $-7.2$ pixels in $y$ to match the sGOODS WCS\null.  Astrometric consistency between surveys is critical for us to accurately match individual LAE components.

\section{RESULTS}
\label{sec:results}

\subsection{Fixed Aperture Half-Light Radii}
\label{subsec:photre}
Table~\ref{tab:GEMSPhot} contains the {\tt PHOT}-derived
$0\farcs6$-aperture magnitudes ($V^{\rm PHOT}$) and half-light radii
($r_e^{\rm PHOT}$) for all LAEs in the {\it HST} surveys.  In GEMS,
six of the 97 objects have no counterpart in the $0\farcs6$ selection
radius, while another six have no detected components, but do
have $0\farcs6$ aperture fluxes at $>2\,\sigma$ level ($V^{\rm
  PHOT}<28.45$).  In GOODS, only one object (LAE 84) has an aperture flux $<2\, \sigma$, but another five have no SExtractor detections.  All three of the HUDF LAEs have SExtractor detections within $0\farcs6$.  Among the LAEs for which we could determine a centroid, there is no evidence for an offset between the Ly$\alpha$ emission and that of the continuum.  The best-fit two-dimensional gaussian to the distribution of these offsets has $\sigma = 0\farcs21$, which is consistent with the expected $\sim 0\farcs2 - 0\farcs3$ astrometric uncertainties of the ground-based observations \citep{MUSYC}.

\begin{figure}[t]
\plotone{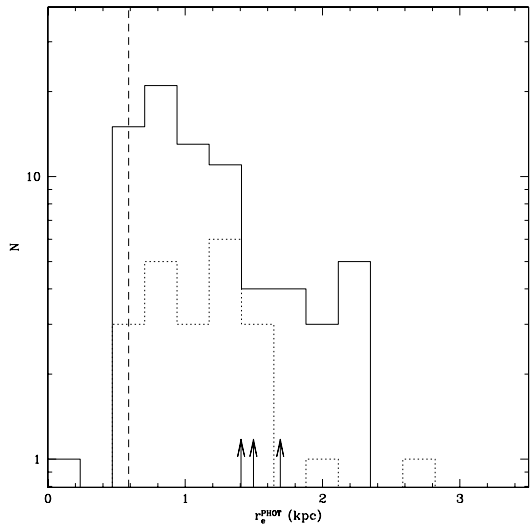}
\caption{Distributions of fixed-aperture, observed half-light radii for objects in GEMS (solid curve), GOODS (dotted), and HUDF (one for each arrow).  The dashed line is the approximate resolution limit of the V-band {\it HST} images.
\label{fig:RePhotHist}}
\end{figure}

Figure~\ref{fig:RePhotHist} displays a histogram of the {\it observed\/} half-light radii for the LAEs in the GEMS (solid), GOODS (dotted), and HUDF (one for each arrow) surveys.  There is a clear excess of sources near the $\sim 0.6$~kpc resolution limit of GEMS and GOODS, suggesting that a typical LAE is either unresolved or only barely resolved at {\it HST} resolution.  The mean half-light radii of the detected LAEs are $\bar{r}_e^{\rm PHOT}=0.98$~kpc, $0.91$~kpc, and $1.53$~kpc in GEMS, GOODS, and HUDF, respectively.  For comparison, \citet{Overzier08} give $\bar{r}_e^{\rm PHOT}=0.9$~kpc as the mean rest-frame UV half-light radii of 12 LAEs at $z=4.1$.

\begin{figure}[t]
\plotone{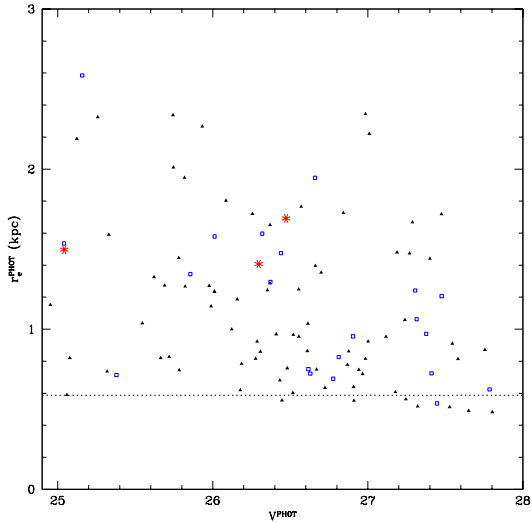}
\caption{Fixed-aperture, rest-UV half-light radius plotted versus
rest-UV continuum magnitude in the full
sample of LAEs with SExtractor detections, including objects in GEMS (black triangles), GOODS
(blue open squares), and HUDF (red asterisks).  The dotted line
indicates the approximate resolution limit of the V-band {\it HST} images.
\label{fig:RpvV}}
\end{figure}

\begin{figure}[t]
\plotone{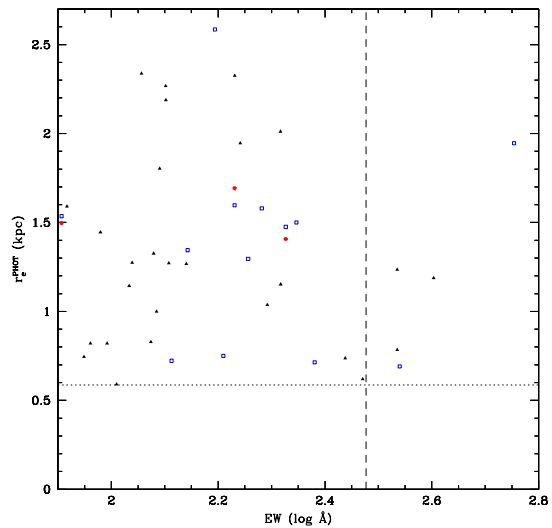}
\caption{ Fixed-aperture, rest-UV half-light radius plotted as a function
of Ly$\alpha$ equivalent width, where EW(Ly$\alpha$) measurements are taken from \citet{GronwallLAE}.  LAEs with S/N~$>30$ in the V-band images are plotted, including objects in GEMS (black triangles), GOODS
(blue open squares), and HUDF (red asterisks).  The dotted line
indicates the approximate resolution limit and the dashed line indicates the
approximate EW(Ly$\alpha$) at which Ly$\alpha$ emission is
$50$\% of the light detected in the V-band filter.
\label{fig:LyaV}}
\end{figure}

Figure~\ref{fig:RpvV} plots the dependence of $r_e^{\rm PHOT}$ with
$V^{\rm PHOT}$.  The GEMS data show little correlation between the two
parameters, but the deeper GOODS data display weak evidence for an
increase in size with increasing flux.  There is also little evidence
for a correlation between the continuum half-light radius, $r_e^{\rm
  PHOT}$, and Ly$\alpha$ equivalent width (EW(Ly$\alpha$), see Figure~\ref{fig:LyaV}).  The EW(Ly$\alpha$) values are estimated in \citet{GronwallLAE} using the broadband and narrow-band photometry.  We plot only LAEs with S/N~$>30$ in Figure~\ref{fig:LyaV} (not that little correlation is seen even when the entire sample is plotted).

\begin{figure}[t]
\plotone{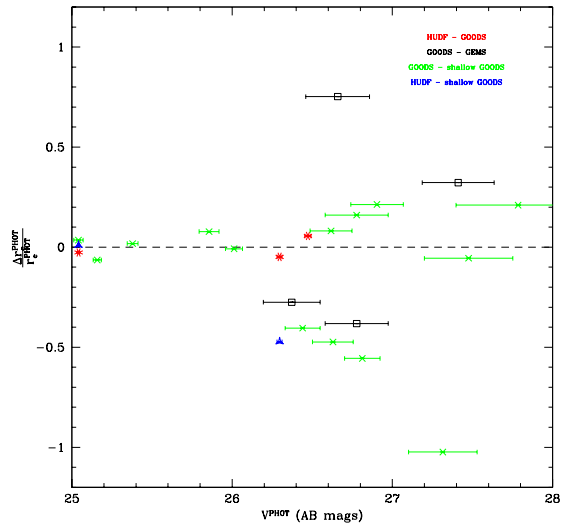}
\caption{ Fractional difference in estimates of the fixed-aperture half-light radii for the same objects
between different surveys, plotted as
a function of V-band magnitude.  For all points, $\Delta r_e^{\rm PHOT}$
indicates the difference between the radius of the deeper minus that of the shallower survey, where green 
crosses are GOODS vs. shallow GOODS, black open squares are GOODS vs. GEMS,
blue triangles are HUDF vs. shallow GOODS, and red asterisks are HUDF
vs. GOODS.  Fixed-aperture measurements of the half-light radius appear
consistent at $V^{\rm PHOT}\lesssim 26.3$, or S/N$\gtrsim 30$ in GEMS.
\label{fig:dRpV}}
\end{figure}

Our curve-of-growth measurement of $r_e^{\rm PHOT}$ is model-independent;
thus, if the rest-UV LAE centroids are well determined, $r_e^{\rm PHOT}$
should be insensitive to the depth of the survey.  However, if the
rest-UV continuum is diffuse or clumpy, the SExtractor detections, and
therefore the derived centroids and $r_e^{\rm PHOT}$ may vary
with depth.  In fact, there is clear evidence for this in
Figure~\ref{fig:dRpV}, which plots the fractional difference in
$r_e^{\rm PHOT}$ between surveys as a function of V-band magnitude.  At
magnitudes brighter than $V^{\rm PHOT} \sim 26.3$ (S/N$\gtrsim 30$ in
GEMS), the half-light radius is robust to $<10$\% in all surveys.
However, only the HUDF-GOODS comparison fares well between $26 <
V^{\rm PHOT} < 27$, and at fainter magnitudes, there is evidence that the
shallower surveys are systematically overestimating the half-light
radius.  Over the entire magnitude range, the variance of $\Delta
r_e^{\rm PHOT}/r_e^{\rm PHOT}$ between GOODS and sGOODS is $\sim 20$\%.  GEMS
is deeper than sGOODS, so if we assume that the GOODS morphological
parameters are accurate, then this would be a conservative estimate of
the average error in $r_e^{\rm PHOT}$ in the GEMS data.

\subsection{SExtractor Results}
\label{subsec:sex}


Of the $97$ LAEs covered by the GEMS survey, $76$ have at least one
component detected within the $0\farcs6$ selection circle, $16$ have
at least two components, and $4$ have at least three components.  For
comparison, \citet{Taniguchi09} find only $2/47$ multi-component LAEs
at $z=5.7$.  While it is tempting to interpret this as evolution in
the number of components, the {\it HST} images used by
\citet{Taniguchi09} are effectively 2.5 magnitudes shallower than even
GEMS, and inspection of our images implies that very few LAEs would be
seen to have multiple components at that depth.

The cutouts for all LAEs in GEMS are plotted in
Figure~\ref{fig:GEMSPanels}, with the components marked by red arrows
and the selection circle shown in black.  Of the objects with multiple
components, eight appear to have a clumpy morphology and may
correspond to merging galaxies or individual star-forming clumps in a
single galaxy.  The remaining eight cutouts have one cleanly defined
detection, and ``fuzz'' that appears just above the noise or as an
extension to the primary source.  We note that, based on the
field density of objects, we expect $\sim 11$~of the components
detected by SExtractor to be unrelated to observed Ly$\alpha$ emission
(see \S~\ref{subsec:cutouts}).  Hence, several of the apparently
clumpy or fuzzy objects shown here may be interlopers.

In Figure~\ref{fig:GOODSPanels}, we plot the $29$ LAEs covered by the
GOODS survey.  Of these, $23$ have at least one component, four have
two components, and one (LAE 4) has five components.  The ground-based
narrow-band magnitude of LAE 4 is the second brightest of the LAEs in
our full sample of $155$ objects, so its complex and extended
morphology may suggest a protocluster or a massive galaxy in the act
of formation.  The rightmost of the two components in LAE 11 may be an
interloper (we expect $\sim3$ contaminants in the GOODS components
sample) due to its large extent and position on the edge of the
selection circle.  Of the remaining three multi-component objects, LAE
25 and LAE 44 appear to be clumpy and LAE 55 is noisy and may be a
single extended object.  Several of the single-component LAEs, such as
LAE 56, 59, and 125, have asymmetric diffuse emission about the
emission centroid; while this is consistent with possible merger activity,
it could also be caused by an asymmetric distribution of diffuse
star formation or dust.  Finally, in Figure~\ref{fig:HUDFPanels}, we
plot cutouts for the three LAEs with HUDF coverage.  All are detected
as a single component within the selection circle and all show
evidence for asymmetric, extended emission in both GOODS and
HUDF.  It is worth noting that these three objects are not necessarily
representative of the overall LAE population; that is, none of the
point-like or faint LAEs seen in GOODS are covered by HUDF.

\begin{figure}[t]
\plotone{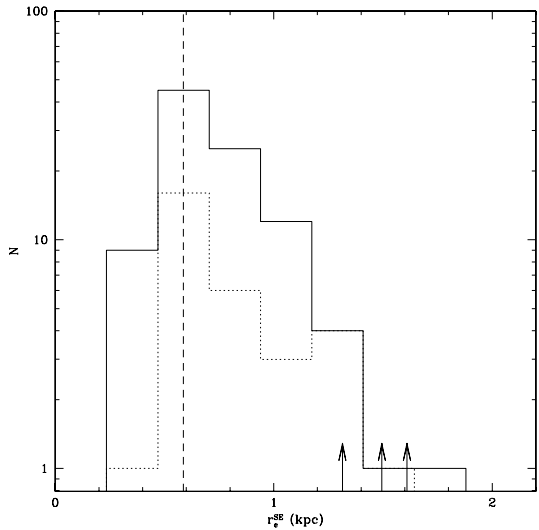}
\caption{Distributions of observed half-light radii, as measured by SExtractor, for LAE components in GEMS (solid curve), GOODS (dotted), and HUDF (one object for each arrow).  The dashed line is the approximate resolution limit of the V-band {\it HST} images.
\label{fig:ReSexHist}}
\end{figure}

The position, brightness, ellipticity, positional angle and observed half-light radius ($r_e^{\rm SE}$) of each LAE component (as computed by SExtractor) are given in Table~\ref{tab:GEMSsex}.  In addition, the $r_e^{\rm SE}$ distributions are given in Figure~\ref{fig:ReSexHist}.  The mean $r_e^{\rm SE}$ of the entire sample of LAE components is $0.74$~kpc in GEMS ($0.79$~kpc in GOODS), while for sources with only one SExtractor detection (i.e., non-clumpy sources) the mean is $0.73$~kpc in GEMS ($0.67$~kpc in GOODS).  This is somewhat smaller than the median size of $r_h \sim 1$~kpc found for non-clumpy $z=3.1$ LAEs by \citet{Venemans05}, but their decision to include only sources with $15$ connected pixels above a $1 \,\sigma$ threshold would have made them insensitive to some of the smaller and fainter objects found in our sample.  Considering this difference in selection criteria, as well as the small number of objects involved (they computed the median half-light radius using only 13 objects), the two results are probably consistent.  In HUDF, all three of the LAEs have a single SExtractor detection within $0\farcs6$, with $\bar{r}_e^{\rm
  SE} = 1.47$~kpc.  Although these deeper observations may be picking
up diffuse emission that is increasing the mean half-light radius, the
sample is too small to draw strong conclusions.

The ability of SExtractor to detect LAE components will clearly depend on survey depth, with the faintest objects likely to go undetected in the shallowest exposures.  As expected, the fraction of LAEs with no counterpart in the {\it HST} images decreases with depth, with $27$\%~($6/22$) in sGOODS, $22$\%~($21/97$) in GEMS, $20$\%~($6/29$) in GOODS, and $0$\%~($0/3$) in HUDF.  Moreover, of the six LAEs not detected in sGOODS, three are present in the full GOODS survey, but all are faint and indistinguishable from point sources (see below).  Finally, we note that in the shallow surveys diffuse emission can go undetected below the sky noise, and a source with a single component in deeper images can be split into multiple components in shallower ones.  This occurs in two of the LAEs (LAE 11 and LAE 125) in the sGOODS survey, but in both cases the vast majority of the total flux is contained in one component.

\begin{figure}[t]
\plotone{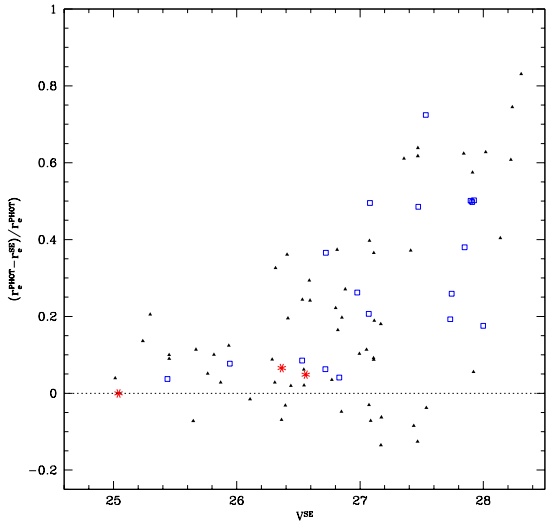}
\caption{Fractional difference between the fixed-aperture half-light radii (computed within $0\farcs6$ apertures) and the SExtractor half-light radii for LAEs with only one detected component, plotted as a function of the rest-UV continuum
magnitude computed by SExtractor.  Objects in GEMS are plotted as black triangles, objects in GOODS as
blue open squares, and objects in HUDF as red asterisks.  The two measures of observed half-light radius appear consistent with one another at $V^{\rm PHOT}\lesssim 26.3$, or S/N$\gtrsim 30$ in GEMS and GOODS.  The dotted line marks $r_e^{\rm PHOT}-r_e^{\rm SE}=0$.
\label{fig:CompareRads}}
\end{figure}

For our chosen set of parameters, SExtractor performs {\tt AUTO}
photometry within an elliptical aperture with radius $2.5R_{\rm Kron}$
\citep{Kron80}, in which $R_{\rm Kron}$ is the first-order moment of
the light distribution.  The parameter, $R_{Kron}$, is in turn
dependent on the radius at which the source flux drops below the
noise.  Since this latter quantity is depth-dependent, we expect
SExtractor to underestimate the half-light radii of faint sources,
particularly those with diffuse emission.  In
Figure~\ref{fig:CompareRads}, we plot the fractional difference
between the {\tt PHOT} half-light radii (computed using our
curve-of-growth analysis; see \S~\ref{subsec:aperture}) and the
SExtractor half-light radii for LAEs with only one detected component.
For LAEs in GEMS with S/N~$\gtrsim 30$ ($V^{\rm SE}
\lesssim 26.3$), the two radii agree to $\sim 10$\%, but then they
diverge rapidly at fainter magnitudes.  The same is true for LAEs in
the GOODS survey, where S/N~$\gtrsim 30$ corresponds to
$V^{\rm SE} \lesssim 26.8$.  We don't have enough objects in HUDF to
determine the flux at which the two radii diverge, but the
half-light radius measurements appear consistent in the three $V^{\rm SE}>26.6$ LAEs
present in the survey.

\subsection{Point Source Samples}
\label{subsec:psres}

\begin{figure}[t]
\plotone{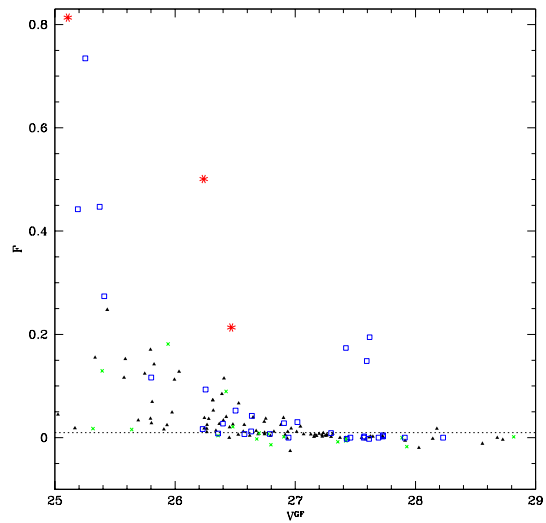}
\caption{Distribution of $F$ (see Equation~\ref{eq:deltachi}) values as a
function of the SExtractor V-band magnitude for LAE components in GEMS (black triangles), GOODS
(blue open squares), shallow GOODS (green crosses), and HUDF (red asterisks).  We
mark with a dotted line $F=F_{\rm crit}=0.01$ -- all of the LAE components appear to be above
this line (and therefore, resolved) at $V^{\rm PHOT}\lesssim 26.3$, or S/N$\gtrsim 30$ in GEMS.
\label{fig:Fdist}}
\end{figure}

Figure~\ref{fig:Fdist} plots the distribution of $F$ (see
Equation~(\ref{eq:deltachi})) as a function of the best-fit V-band
magnitude calculated by {\tt GALFIT} ($V^{\rm GF}$).  LAE components
imaged in the GEMS and GOODS surveys (shown as black triangles and
blue open squares, respectively) are consistently resolved at $V^{\rm
  GF}\lesssim 26.5$ and consistently unresolved at $V^{\rm GF}\gtrsim
27$.  That the dimmest components are consistent with a point source
is simply a reflection of the fact that objects barely detected above
the sky noise can be fit just as well with a three-parameter PSF as
with the seven-parameter S\'{e}rsic profile.  There are three
anomalous components in the GOODS sample at $V^{\rm GF}\sim 27.5$, all
of which are members of the morphologically complex system, LAE 5, and
appear inconsistent with point sources despite their faint magnitudes.

It is important to note that not all of the ``point source'' LAEs are isolated.  In the GEMS sample, only 20 of the 45 unresolved sources had no other object within the selection circle.  Of the remaining 25 components, 12 appear to be part of a multi-component source, and 13 appear to the eye to be extensions of a larger, amorphous object that was split by SExtractor.  As discussed in \S~\ref{subsec:cutouts}, we expect $\sim 11$ contaminants in our sample, so some of these components must be chance superpositions and not associated with the Ly$\alpha$ emission.  In GOODS, a larger fraction ($70$\%) of the 23 unresolved sources are isolated, perhaps due to the decreased tendency for LAEs to be split into multiple components (see \ref{subsec:sex}).

Since the brightest LAE components are all resolved, it is possible that
a deeper survey would resolve many of the apparent point sources.  
Indeed, the fraction of unresolved LAE components drops from $63$\%~($12/17$)
in sGOODS to $47$\% (45/95) in GEMS and $48$\%~(15/31) in the full 
GOODS survey.  Moreover, only four of the 12 point sources in sGOODS 
remain unresolved at GOODS depth.  In the HUDF, only one of the three
sources is consistent with a point source, and it appears as an 
extension to a brighter, resolved component.

\section{DISCUSSION}
\label{sec:discussion}

\subsection{Optimal Techniques for LAE Morphological Analysis}
\label{subsec:discussmethod}

When analyzing the morphologies of LAEs, there is a great deal of ambiguity as to how one should treat objects with multiple clumps.  Some components may be the result of merging/interacting galaxies; others may simply be individual star-formation regions within a single system.  In this paper, we have considered both possibilities, presenting magnitudes and half-light radii both for the LAE system as a whole (using fixed apertures about the light centroid) and for individual LAE components.  We find that the majority of multi-component LAEs identified in shallow frames become connected on deeper images.  This suggests that the majority of rest-UV ``clumps'' are actually individual star-forming regions within a single system.

The presence of this diffuse emission connecting individual clumps suggests that, in the absence of interlopers, LAE radii and total magnitudes should be determined using fixed aperture measurements.  SExtractor-like adaptive techniques that rely on isophotal radii will tend to underestimate an LAE's true half-light radius since they only consider the extended emission surrounding the brightest clumps (see Figure~\ref{fig:dRpV}).  The distinction between definitions of half-light radius is particularly important when comparing the half-light radii of LAEs to those of the more extended Lyman Alpha Blobs \citep[hereafter, LABs,][]{FirstBlobs, FirstBlobs2}, which exhibit a wide range of Ly$\alpha$ morphologies and may be the precursors to present-day rich-cluster galaxies \citep{Yang09}.  Similarly, higher-order non-parametric morphological diagnostics, such as CAS \citep[concetration, asymmetry, and clumpiness,][]{CAS} or the Gini coefficient, should also be performed within a fixed aperture, and not be tied to isophotal radii.  Conversely, parametric profile fits that do not account for a clumpy light distribution are best performed on individual components, as these are more closely approximated by a smooth light distribution.

An additional concern involves the reliability of morphological analyses in different signal-to-noise regimes.  Figures~\ref{fig:RpvV}, \ref{fig:dRpV}, and \ref{fig:Fdist} suggest that one needs a S/N of at least $\sim 30$ in the compact core of the LAE in order to resolve the rest-UV continuum and to make a reasonable estimate of its half-light radius.  For the fixed-aperture measurements, the primary limitation is finding the LAE system centroid; for isophotal measurements, it is image classification, and whether the separate components on shallow frames are actually brighter knots of a larger object.

\subsection{LAE Sizes and Morphologies}
\label{subsec:discussLAE}

The results presented in \S~\ref{sec:results} suggest that LAEs at $z \sim 3$ are generally $\lesssim 2$~kpc in size in the rest-frame UV, while the individual components of an LAE system are typically $\lesssim 1.5$~kpc.  Both of these results are consistent with previous work.  \citet{Gawiser07} has shown that the majority of LAEs are likely to be in the early phases of a starburst, perhaps even experiencing their first large-scale burst of star-formation.  Consequently, we don't expect their sizes or morphologies to vary greatly with wavelength.  Even the more massive, and presumably older, LBGs have been shown to have a negligible morphological $k$-correction between the observed-frame optical and near-infrared \citep{Dickinson}.  However, we should {\it not\/} use the rest-UV morphology to infer the extent and distribution of the Ly$\alpha$ emission.  At $z=3.1$, the V-band probes the rest-UV continuum light from star-forming regions associated with the host galaxy of the LAE\null.  At low redshift, most of the Ly$\alpha$ emission originates in a diffuse halo surrounding the galaxy \citep{Ostlin08} .  Presumably, this is a consequence of resonant scattering in the Ly$\alpha$ line; if the same process occurs at high redshift, then an LAE's Ly$\alpha$-emission-line morphology will be ``smeared'' relative to the distribution of its star-forming regions.

If there were extended Ly$\alpha$ halos in a large fraction of
LAEs, we might expect to see a correlation between the V-band
half-light radius and the equivalent width of the Ly$\alpha$ line due
to the increased contribution of the extended Ly$\alpha$ emission to
the V-band flux.  However, this effect would only begin to appear at
EW(Ly$\alpha$)$\gtrsim300$\AA\ (marked by a dashed line in
Figure~\ref{fig:LyaV}), above which $\gtrsim50$\% of the V-band light
comes from the emission line.  There are only five objects in our
sample that meet this criterion and, given the range of half-light
radii seen at smaller EW(Ly$\alpha$), it is impossible to say anything
about the existence or extent of Ly$\alpha$ halos from this sample
alone.

A more direct method of searching for Ly$\alpha$ halos would be to
observe LAEs at high resolution in a narrow-band filter.  There are
currently no published studies of LAE morphologies in Ly$\alpha$
emission, but one is in progress for a subset of the current MUSYC
sample (Gronwall 2009, in preparation).  Preliminary results from an
ACS narrow-band survey of LAEs (Bram Venemans, private communication)
suggest that high$-z$ LAEs do indeed have Ly$\alpha$ halos, as the
Ly$\alpha$ emission detected in high-resolution images often cannot
account for all the flux seen from the ground.  Moreover, even in
ground-based images, there is evidence that $z \sim 2$ LAEs are more
extended in the emission line than in the continuum
\citep{NilssonLAE}.

\subsection{Star Formation in LAEs}
\label{subsec:discussSFR}

As shown in \citet{Gawiser07}, very few of the LAEs in our sample are detected at X-Ray wavelengths and there is no evidence for high-ionization emission lines in the rest-UV spectra of the remaining objects for which we have spectral information.  This suggests that AGN are unlikely to be the power source for the Ly$\alpha$ emission.  Although a low-luminosity or obscured AGN may be present in some of these sources \citep{Finkelstein09}, the fact that the rest-UV light distribution is consistently resolved at $S/N \gtrsim 30$ (see Figure~\ref{fig:Fdist}) suggests that any ionizing flux is likely coming from massive stars rather than a nuclear source.  In addition, the correlation between UV- and Ly$\alpha$-based estimates of the star formation rate seen in \citet{GronwallLAE} suggests that shock ionization is also not a substantial source of power for the line emission.

In the last column of Table~\ref{tab:GEMSsex}, we give the SFRs for individual LAE components, estimated using their rest-frame UV flux (given by $V^{SE}$) and
the standard conversion \citep[][see their equation 1]{Kennicutt98}, assuming a
Salpeter IMF and a negligible dust correction.  The SFRs for LAE components range from $\sim 0.1$~M$_{\sun}$~yr$^{-1}$ to $\sim 5$~M$_{\sun}$~yr$^{-1}$.  The
sum of SFRs in individual components is within $10-20$\% of the SFR for the composite system (as inferred from $V^{PHOT}$) when S/N~$\gtrsim 30$ for the system.
  This is consistent with the difference between the half-light radii determined
 with {\tt PHOT} and SExtractor for single-component systems (see Figure~\ref{fig:CompareRads}).  In addition, we find that $8/15$ of the two-component objects
have SFR ratios less than 3:1.  Although this could be interpreted as evidence for major merger events
between individual components, the high rate of contamination expected in two-component objects and the depth dependence of the component segregation make it difficult to determine which, if any, of these LAEs are ongoing major mergers.

Considering that LAEs are thought to have stellar masses of $M \sim 10^9$~M$_{\sun}$ \citep{Gawiser07}, there is no local analog for this level of star formation activity in objects of comparable mass.  However, the SFRs and sizes of LAE components ($\lesssim 1$~kpc) are comparable to those of the nucleii of M82-like starburst galaxies in the local universe \citep{Mayya04}.  At $z\sim 3$, galaxies identified using other selection techniques, such as LBGs and submillimeter galaxies, have typical star formation rates that are at least an order of magnitude larger than those in LAEs and their components \citep{Shapley01,Genzel03}.  However, LBGs have also been shown to exhibit clumpy star formation \citep{Papovich05} and may be undergoing a dynamical process similar to that leading tothe active star formation and line emission seen in LAEs.  An application of the
 pipeline developed here to LBGs would help to elucidate this comparison.


\acknowledgments

Support for this work was provided by NASA through grant number
HST-AR-11253.01-A from the Space Telescope Science Institute, which is
operated by AURA, Inc., under NASA contract NAS 5-26555 and by the
National Science Foundation under grant AST-0807570.  C. Gronwall
was provided support through grant number HST-AR-10324.01-A.  We thank Peter
Kurczynski for his helpful suggestions on the point source analysis.

\clearpage
\bibliographystyle{apj}                       

\bibliography{apj-jour,Bond0713}    

\onecolumn

\newpage
\begin{figure*}[t]
\plotone{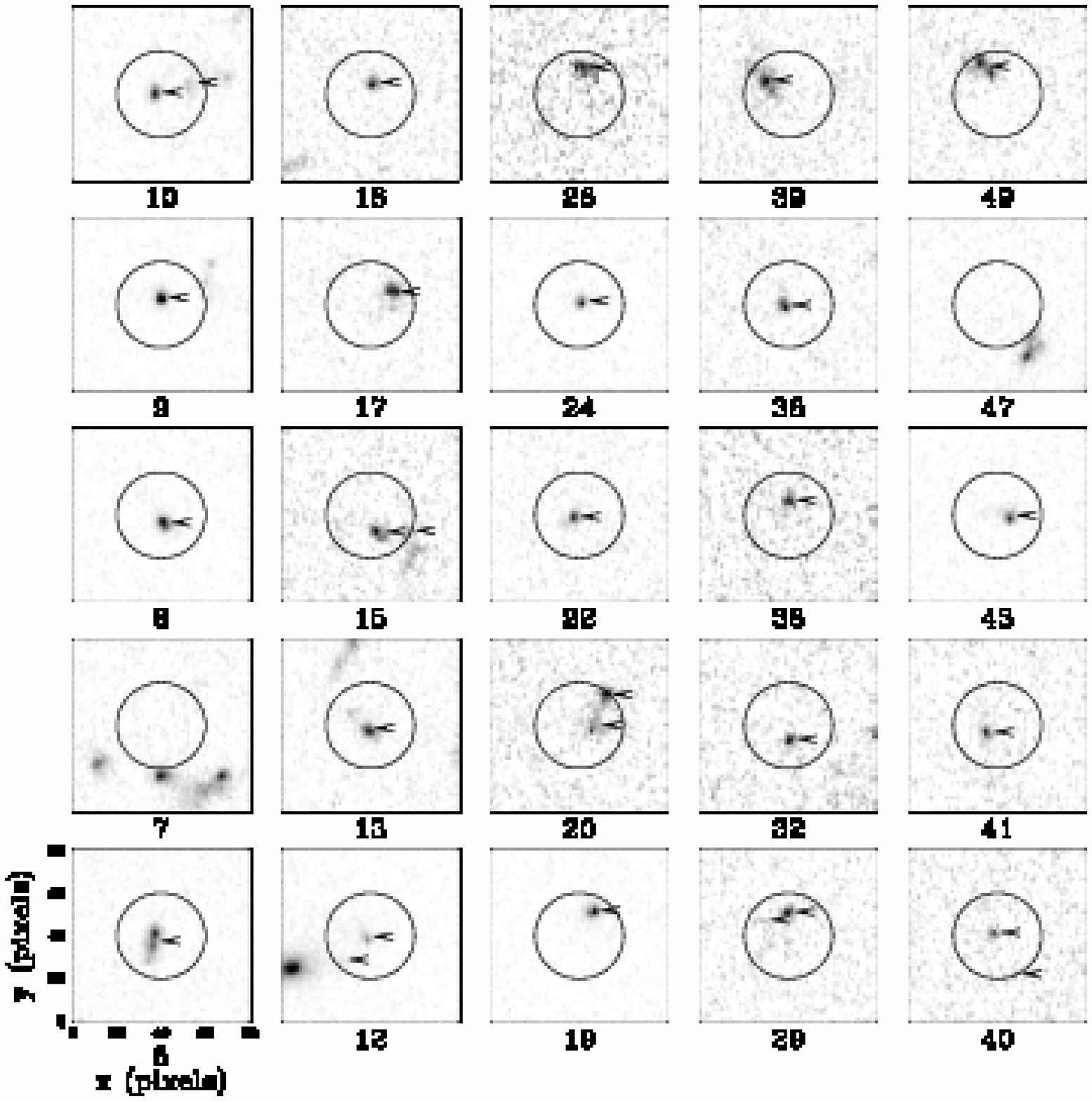}
\caption{LAE cutouts extracted from
the GEMS survey images.  We mark components with red arrows and
draw the selection circle in black.  Numbers underneath the panels are
the corresponding LAE indeces from \citet{GronwallLAE}.
\label{fig:GEMSPanels}}
\end{figure*}

\clearpage
\newpage
\begin{figure*}[t]
\figurenum{10}
\plotone{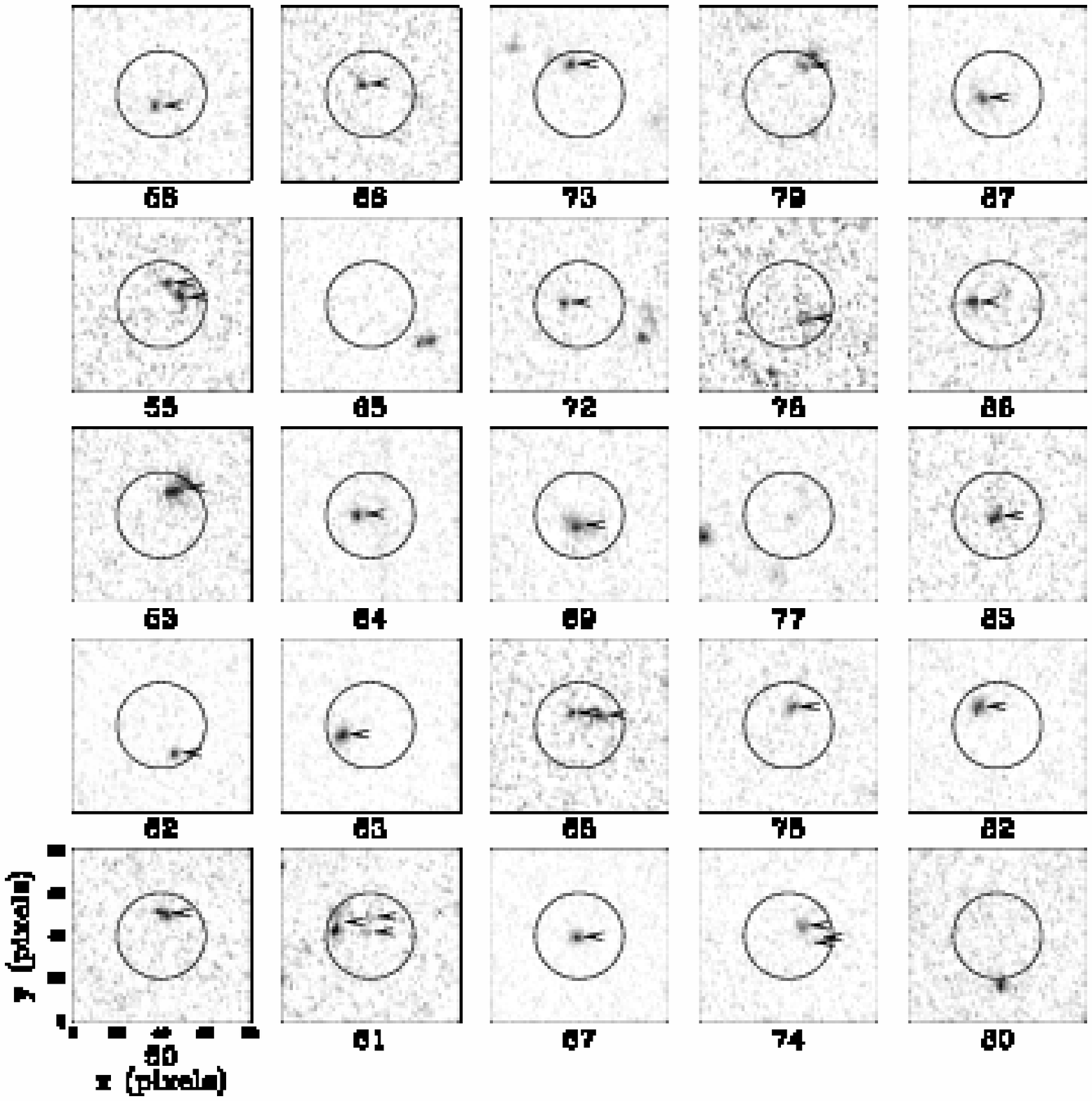}
\caption{(cont.)}
\end{figure*}

\newpage
\begin{figure*}[t]
\figurenum{10}
\plotone{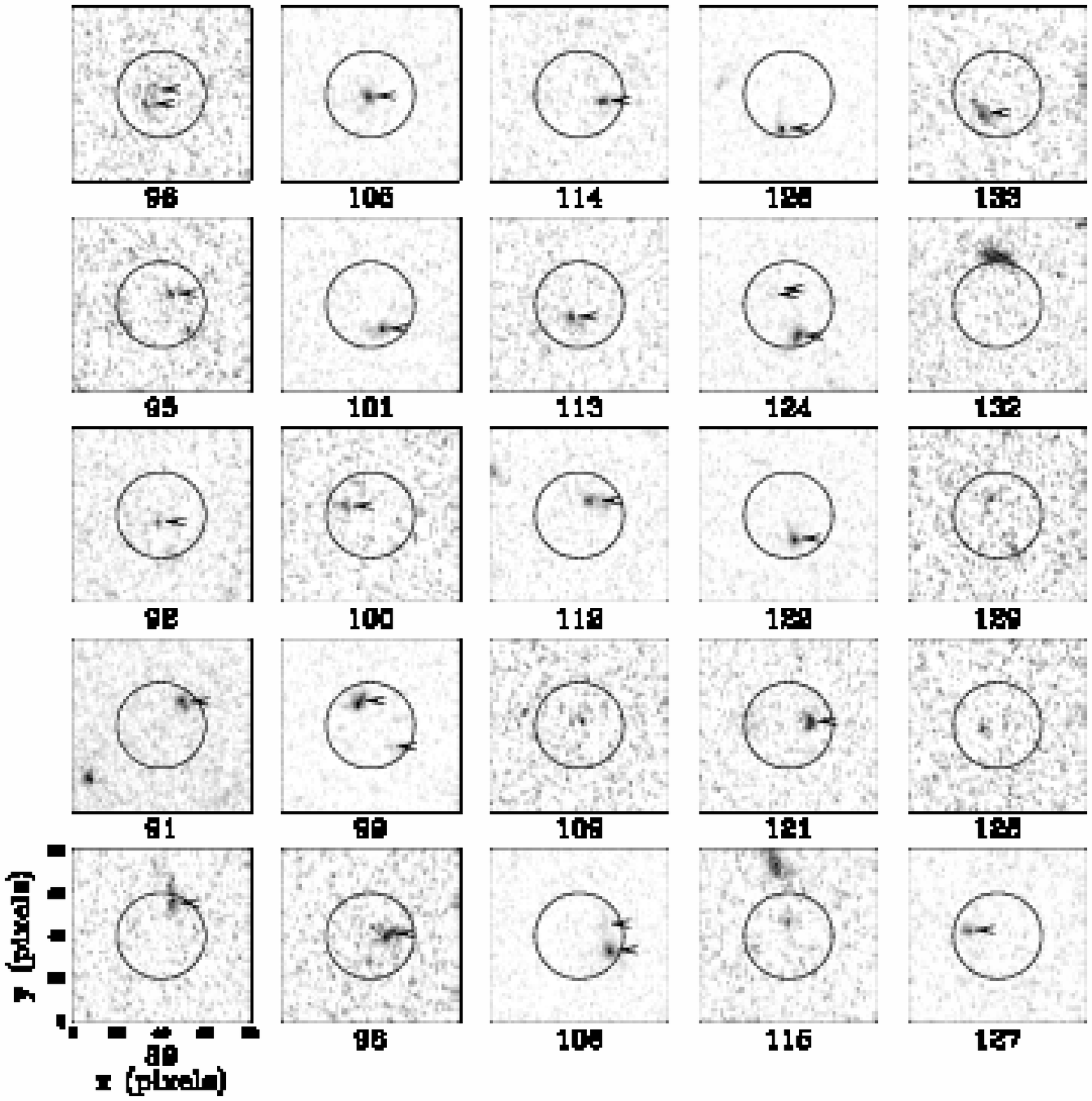}
\caption{(cont.)}
\end{figure*}

\begin{figure*}[t]
\figurenum{10}
\plotone{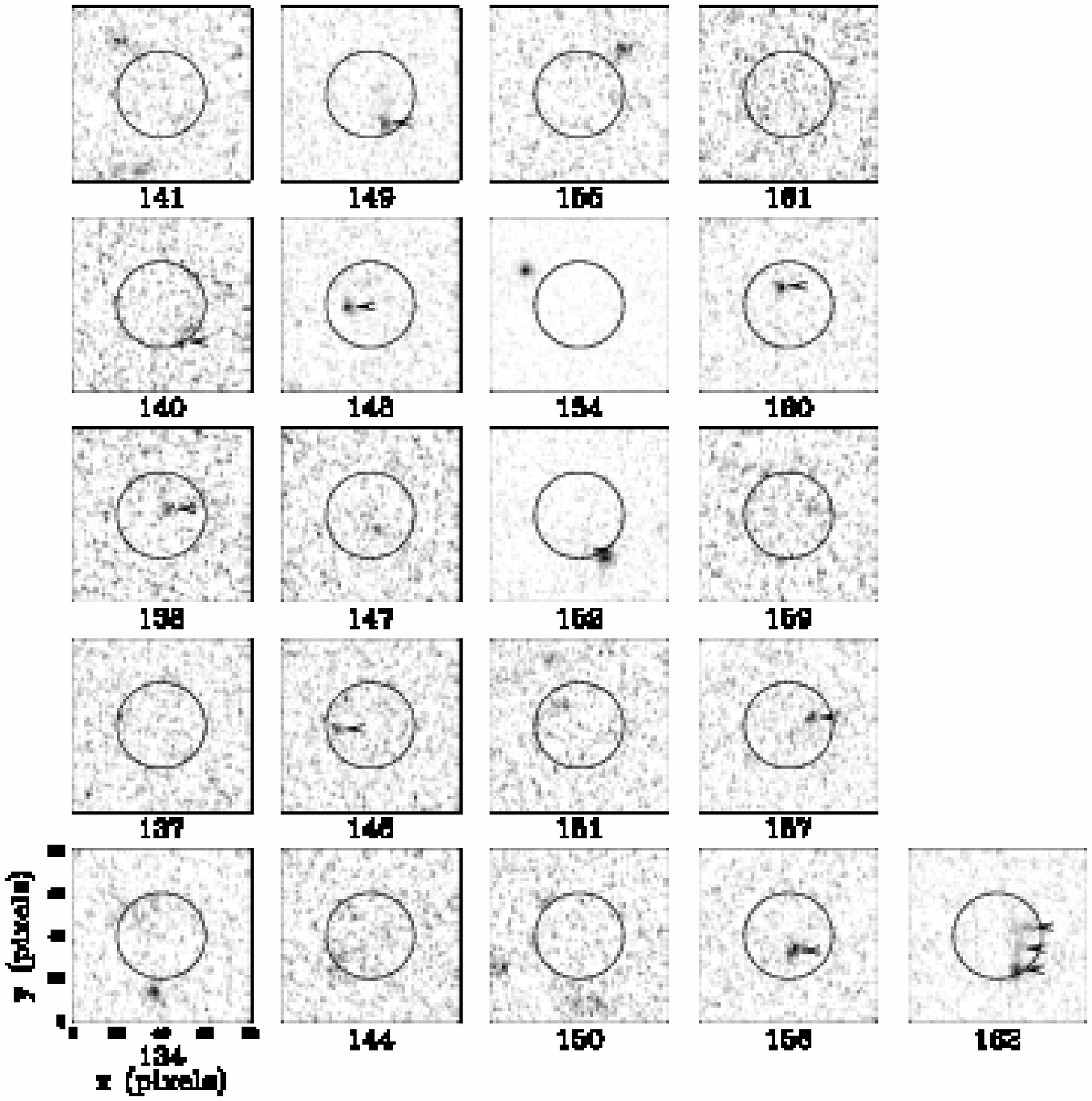}
\caption{(cont.)}
\end{figure*}

\begin{figure*}[t]
\plotone{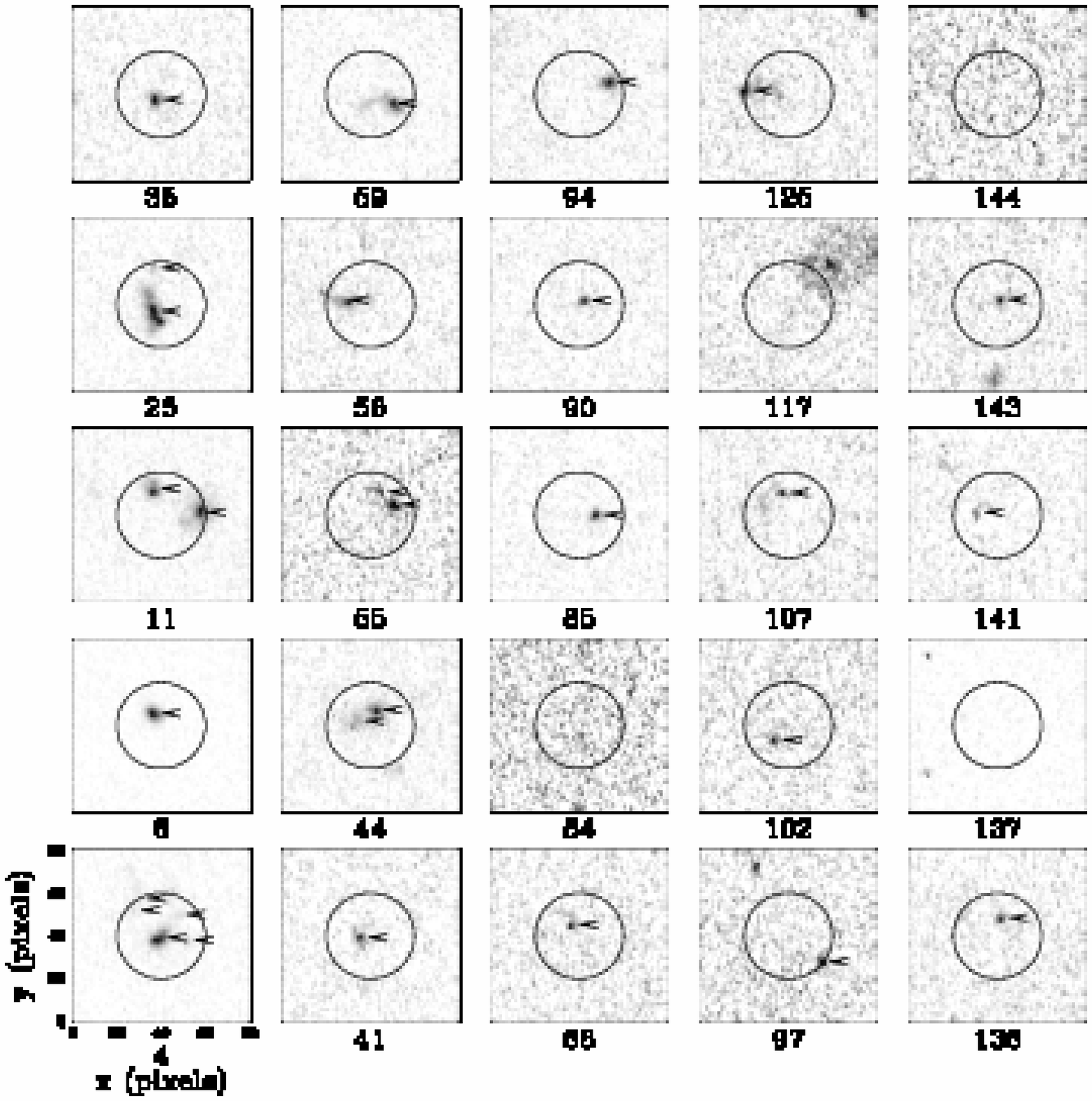}
\caption{Same as Figure~\ref{fig:GEMSPanels}, but for LAEs detected in the GOODS survey.
\label{fig:GOODSPanels}}
\end{figure*}

\begin{figure*}[t]
\figurenum{11}
\plotone{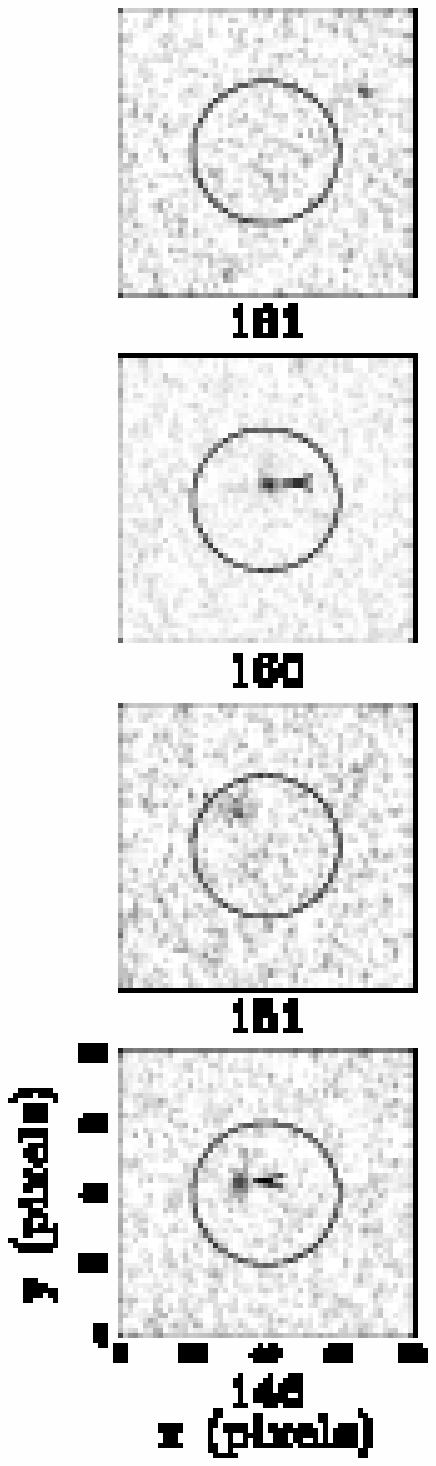}
\caption{(cont.)}
\end{figure*}

\begin{figure*}[t]
\plotone{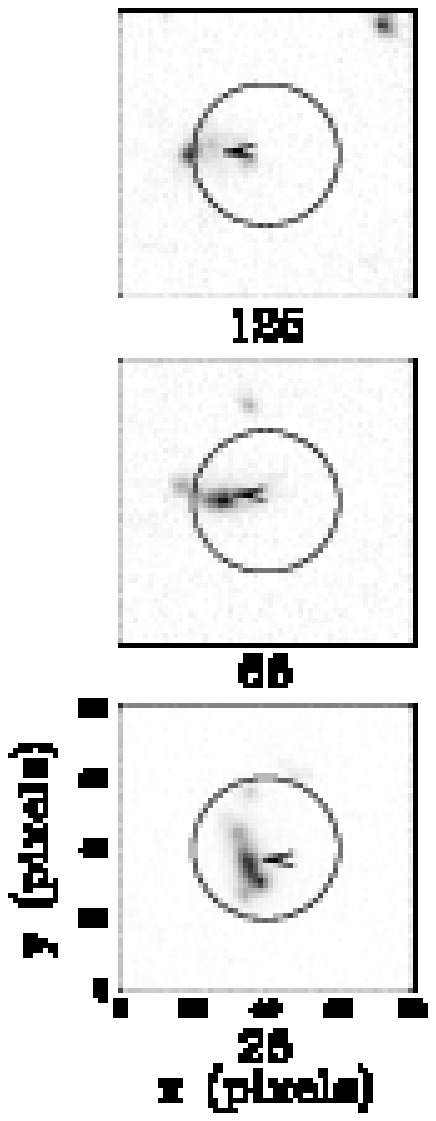}
\caption{Same as Figure~\ref{fig:GEMSPanels}, but for LAEs detected in the HUDF survey.
\label{fig:HUDFPanels}}
\end{figure*}

\begin{deluxetable}{lcccccc}
\tablecaption{LAE Photometric Properties in GEMS\label{tab:GEMSPhot}}
\tablewidth{0pt}
\tablehead{
\colhead{Number\tablenotemark{a}}
&\colhead{Survey}
&\colhead{$\alpha$\tablenotemark{b}}
&\colhead{$\delta$\tablenotemark{b}}
&\colhead{$V^{\rm PHOT}$}
&\colhead{$d_c$\tablenotemark{c}}
&\colhead{$r_e^{\rm PHOT}$~\tablenotemark{d}}
\\
&
&
&
&\colhead{(AB mags)}
&\colhead{(\arcsec)}
&\colhead{(\arcsec)}
}
\startdata
$25$ &HUDF &$3$:$32$:$40.785$ &$-27$:$46$:$06.037$ &$25.04 \pm 0.01$ &$0.19$ &$0.19$  \\
$56$ &HUDF &$3$:$32$:$34.328$ &$-27$:$47$:$59.545$ &$26.30 \pm 0.02$ &$0.43$ &$0.18$  \\
$125$ &HUDF &$3$:$32$:$39.013$ &$-27$:$46$:$22.311$ &$26.47 \pm 0.02$ &$0.53$ &$0.22$  \\
$4$ &GOODS &$3$:$32$:$18.813$ &$-27$:$42$:$48.103$ &$24.89 \pm 0.03$ &$0.36$ &$0.19$  \\
$6$ &GOODS &$3$:$32$:$52.690$ &$-27$:$48$:$09.284$ &$25.38 \pm 0.03$ &$0.19$ &$0.09$  \\
$11$ &GOODS &$3$:$32$:$26.937$ &$-27$:$41$:$27.937$ &$25.16 \pm 0.03$ &$0.33$ &$0.33$  \\
$25$ &GOODS &$3$:$32$:$40.785$ &$-27$:$46$:$05.997$ &$25.04 \pm 0.03$ &$0.16$ &$0.20$  \\
$35$ &GOODS &$3$:$32$:$45.604$ &$-27$:$52$:$10.914$ &$26.81 \pm 0.11$ &$0.16$ &$0.11$  \\
$41$ &GOODS &$3$:$32$:$56.672$ &$-27$:$49$:$49.199$ &$26.78 \pm 0.20$ &$0.19$ &$0.09$  \\
$44$ &GOODS &$3$:$32$:$15.799$ &$-27$:$44$:$09.993$ &$26.01 \pm 0.05$ &$0.45$ &$0.20$  \\
$55$ &GOODS &$3$:$32$:$59.976$ &$-27$:$50$:$26.308$ &$26.37 \pm 0.18$ &$0.31$ &$0.17$  \\
$56$ &GOODS &$3$:$32$:$34.331$ &$-27$:$47$:$59.554$ &$26.44 \pm 0.11$ &$0.46$ &$0.19$  \\
$59$ &GOODS &$3$:$32$:$33.254$ &$-27$:$51$:$27.590$ &$25.86 \pm 0.06$ &$0.32$ &$0.17$  \\
$66$ &GOODS &$3$:$32$:$48.528$ &$-27$:$56$:$05.376$ &$26.66 \pm 0.20$ &$0.19$ &$0.25$  \\
$84$ &GOODS &$3$:$32$:$17.395$ &$-27$:$42$:$48.440$ &$29.07 \pm 1.20$ &$...$ &$...$  \\
$85$ &GOODS &$3$:$32$:$59.824$ &$-27$:$53$:$05.768$ &$26.62 \pm 0.13$ &$0.23$ &$0.10$  \\
$90$ &GOODS &$3$:$32$:$14.574$ &$-27$:$45$:$52.417$ &$26.90 \pm 0.16$ &$0.30$ &$0.12$  \\
$94$ &GOODS &$3$:$32$:$09.336$ &$-27$:$43$:$54.193$ &$26.63 \pm 0.13$ &$0.40$ &$0.09$  \\
$97$ &GOODS &$3$:$32$:$12.584$ &$-27$:$48$:$05.665$ &$27.30 \pm 0.21$ &$0.63$ &$0.16$  \\
$102$ &GOODS &$3$:$32$:$57.356$ &$-27$:$51$:$42.357$ &$27.38 \pm 0.29$ &$0.34$ &$0.12$  \\
$107$ &GOODS &$3$:$32$:$10.158$ &$-27$:$44$:$28.336$ &$27.48 \pm 0.28$ &$0.60$ &$0.15$  \\
$117$ &GOODS &$3$:$32$:$12.983$ &$-27$:$44$:$52.140$ &$27.77 \pm 0.35$ &$...$ &$...$  \\
$125$ &GOODS &$3$:$32$:$39.016$ &$-27$:$46$:$22.305$ &$26.32 \pm 0.09$ &$0.58$ &$0.20$  \\
$136$ &GOODS &$3$:$32$:$24.329$ &$-27$:$41$:$51.886$ &$27.32 \pm 0.21$ &$0.47$ &$0.14$  \\
$137$ &GOODS &$3$:$32$:$59.951$ &$-27$:$50$:$29.180$ &$27.63 \pm 0.56$ &$...$ &$...$  \\
$141$ &GOODS &$3$:$33$:$02.097$ &$-27$:$51$:$07.663$ &$27.45 \pm 0.47$ &$0.34$ &$0.07$  \\
$143$ &GOODS &$3$:$32$:$54.652$ &$-27$:$51$:$49.228$ &$27.79 \pm 0.39$ &$0.03$ &$0.08$  \\
$144$ &GOODS &$3$:$32$:$47.581$ &$-27$:$55$:$38.250$ &$28.67 \pm 1.27$ &$...$ &$...$  \\
$146$ &GOODS &$3$:$32$:$44.168$ &$-27$:$50$:$56.842$ &$28.75 \pm 0.88$ &$0.27$ &$0.04$  \\
$151$ &GOODS &$3$:$32$:$49.749$ &$-27$:$46$:$43.793$ &$28.95 \pm 1.26$ &$0.38$ &$0.04$  \\
$160$ &GOODS &$3$:$32$:$02.802$ &$-27$:$45$:$28.708$ &$27.41 \pm 0.22$ &$0.39$ &$0.09$  \\
$161$ &GOODS &$3$:$33$:$02.969$ &$-27$:$50$:$29.420$ &$28.51 \pm 1.32$ &$...$ &$...$  \\
$5$ &GEMS &$3$:$32$:$47.516$ &$-27$:$58$:$07.706$ &$24.95 \pm 0.03$ &$0.18$ &$0.15$  \\
$7$ &GEMS &$3$:$31$:$44.988$ &$-27$:$35$:$32.870$ &$27.45 \pm 0.32$ &$...$ &$...$  \\
$8$ &GEMS &$3$:$31$:$54.885$ &$-27$:$51$:$21.114$ &$25.32 \pm 0.04$ &$0.14$ &$0.09$  \\
$9$ &GEMS &$3$:$31$:$40.157$ &$-28$:$03$:$07.406$ &$25.06 \pm 0.03$ &$0.08$ &$0.08$  \\
$10$ &GEMS &$3$:$33$:$22.442$ &$-27$:$46$:$36.850$ &$25.12 \pm 0.04$ &$0.19$ &$0.28$  \\
$12$ &GEMS &$3$:$32$:$33.846$ &$-27$:$36$:$35.117$ &$25.26 \pm 0.04$ &$0.25$ &$0.30$  \\
$13$ &GEMS &$3$:$33$:$07.252$ &$-27$:$47$:$47.176$ &$25.72 \pm 0.07$ &$0.09$ &$0.11$  \\
$15$ &GEMS &$3$:$33$:$18.915$ &$-27$:$38$:$28.469$ &$25.75 \pm 0.07$ &$0.35$ &$0.26$  \\
$17$ &GEMS &$3$:$32$:$49.148$ &$-27$:$34$:$39.971$ &$25.55 \pm 0.06$ &$0.37$ &$0.13$  \\
$18$ &GEMS &$3$:$32$:$46.754$ &$-27$:$39$:$59.914$ &$26.72 \pm 0.16$ &$0.15$ &$0.08$  \\
$19$ &GEMS &$3$:$31$:$34.738$ &$-27$:$56$:$21.821$ &$25.08 \pm 0.04$ &$0.38$ &$0.10$  \\
$20$ &GEMS &$3$:$33$:$11.879$ &$-28$:$00$:$12.377$ &$25.82 \pm 0.07$ &$0.35$ &$0.25$  \\
$22$ &GEMS &$3$:$31$:$51.636$ &$-27$:$58$:$32.642$ &$26.18 \pm 0.10$ &$0.12$ &$0.10$  \\
$24$ &GEMS &$3$:$31$:$53.213$ &$-27$:$57$:$08.160$ &$26.18 \pm 0.09$ &$0.03$ &$0.08$  \\
$26$ &GEMS &$3$:$31$:$51.561$ &$-27$:$46$:$47.016$ &$26.16 \pm 0.10$ &$0.35$ &$0.15$  \\
$29$ &GEMS &$3$:$31$:$47.800$ &$-27$:$45$:$03.384$ &$26.35 \pm 0.12$ &$0.25$ &$0.16$  \\
$32$ &GEMS &$3$:$33$:$19.793$ &$-27$:$38$:$20.505$ &$27.25 \pm 0.26$ &$0.22$ &$0.07$  \\
$36$ &GEMS &$3$:$32$:$18.926$ &$-27$:$38$:$40.195$ &$26.98 \pm 0.21$ &$0.18$ &$0.10$  \\
$38$ &GEMS &$3$:$31$:$50.369$ &$-27$:$59$:$10.119$ &$26.52 \pm 0.13$ &$0.07$ &$0.08$  \\
$39$ &GEMS &$3$:$31$:$30.524$ &$-27$:$47$:$29.647$ &$25.33 \pm 0.04$ &$0.40$ &$0.20$  \\
$40$ &GEMS &$3$:$31$:$31.471$ &$-27$:$34$:$47.404$ &$27.19 \pm 0.26$ &$0.12$ &$0.19$  \\
$41$ &GEMS &$3$:$32$:$56.672$ &$-27$:$49$:$49.241$ &$26.55 \pm 0.14$ &$0.21$ &$0.12$  \\
$43$ &GEMS &$3$:$33$:$07.315$ &$-27$:$54$:$38.983$ &$25.78 \pm 0.07$ &$0.17$ &$0.10$  \\
$47$ &GEMS &$3$:$32$:$43.499$ &$-27$:$38$:$08.880$ &$27.13 \pm 0.25$ &$...$ &$...$  \\
$49$ &GEMS &$3$:$31$:$42.359$ &$-27$:$58$:$07.857$ &$25.69 \pm 0.06$ &$0.44$ &$0.16$  \\
$50$ &GEMS &$3$:$31$:$52.830$ &$-27$:$45$:$18.650$ &$27.18 \pm 0.24$ &$0.28$ &$0.08$  \\
$52$ &GEMS &$3$:$33$:$21.363$ &$-27$:$38$:$36.337$ &$26.91 \pm 0.19$ &$0.45$ &$0.07$  \\
$53$ &GEMS &$3$:$32$:$15.131$ &$-27$:$38$:$53.946$ &$25.82 \pm 0.07$ &$0.42$ &$0.16$  \\
$55$ &GEMS &$3$:$32$:$59.982$ &$-27$:$50$:$26.367$ &$26.37 \pm 0.12$ &$0.21$ &$0.21$  \\
$58$ &GEMS &$3$:$33$:$06.943$ &$-27$:$42$:$27.852$ &$26.61 \pm 0.14$ &$0.21$ &$0.13$  \\
$61$ &GEMS &$3$:$33$:$09.421$ &$-27$:$45$:$50.114$ &$26.09 \pm 0.07$ &$0.38$ &$0.23$  \\
$63$ &GEMS &$3$:$32$:$51.917$ &$-27$:$42$:$12.246$ &$25.66 \pm 0.06$ &$0.47$ &$0.10$  \\
$64$ &GEMS &$3$:$31$:$59.831$ &$-27$:$49$:$46.424$ &$25.99 \pm 0.08$ &$0.19$ &$0.15$  \\
$65$ &GEMS &$3$:$31$:$42.733$ &$-27$:$53$:$05.433$ &$28.47 \pm 0.81$ &$0.41$ &$0.34$  \\
$66$ &GEMS &$3$:$32$:$48.527$ &$-27$:$56$:$05.357$ &$27.80 \pm 0.43$ &$0.19$ &$0.06$  \\
$67$ &GEMS &$3$:$32$:$51.770$ &$-27$:$37$:$33.552$ &$26.45 \pm 0.12$ &$0.04$ &$0.07$  \\
$68$ &GEMS &$3$:$32$:$58.140$ &$-27$:$48$:$04.877$ &$26.70 \pm 0.16$ &$0.22$ &$0.17$  \\
$69$ &GEMS &$3$:$33$:$25.357$ &$-28$:$02$:$46.531$ &$26.12 \pm 0.09$ &$0.16$ &$0.13$  \\
$72$ &GEMS &$3$:$33$:$07.179$ &$-27$:$48$:$51.544$ &$27.00 \pm 0.21$ &$0.25$ &$0.12$  \\
$73$ &GEMS &$3$:$32$:$57.404$ &$-27$:$55$:$19.073$ &$26.31 \pm 0.11$ &$0.41$ &$0.11$  \\
$74$ &GEMS &$3$:$33$:$18.588$ &$-27$:$45$:$42.617$ &$26.67 \pm 0.13$ &$0.24$ &$0.10$  \\
$75$ &GEMS &$3$:$32$:$59.268$ &$-27$:$41$:$14.756$ &$26.87 \pm 0.19$ &$0.24$ &$0.10$  \\
$77$ &GEMS &$3$:$31$:$54.678$ &$-27$:$52$:$52.927$ &$27.75 \pm 0.42$ &$0.06$ &$0.11$  \\
$78$ &GEMS &$3$:$33$:$20.840$ &$-27$:$51$:$45.883$ &$27.58 \pm 0.35$ &$0.32$ &$0.10$  \\
$79$ &GEMS &$3$:$31$:$58.027$ &$-27$:$47$:$30.335$ &$26.26 \pm 0.09$ &$0.41$ &$0.22$  \\
$80$ &GEMS &$3$:$33$:$20.491$ &$-27$:$36$:$40.140$ &$28.87 \pm 1.21$ &$...$ &$...$  \\
$82$ &GEMS &$3$:$31$:$47.776$ &$-27$:$42$:$16.325$ &$26.43 \pm 0.12$ &$0.37$ &$0.09$  \\
$83$ &GEMS &$3$:$31$:$38.670$ &$-27$:$45$:$43.587$ &$26.52 \pm 0.11$ &$0.08$ &$0.12$  \\
$86$ &GEMS &$3$:$33$:$28.159$ &$-28$:$03$:$20.770$ &$27.27 \pm 0.27$ &$0.24$ &$0.19$  \\
$87$ &GEMS &$3$:$33$:$05.027$ &$-27$:$43$:$37.306$ &$25.98 \pm 0.08$ &$0.24$ &$0.16$  \\
$89$ &GEMS &$3$:$33$:$12.017$ &$-27$:$58$:$39.929$ &$26.66 \pm 0.16$ &$0.45$ &$0.18$  \\
$91$ &GEMS &$3$:$31$:$58.803$ &$-27$:$49$:$28.767$ &$26.91 \pm 0.19$ &$0.44$ &$0.08$  \\
$92$ &GEMS &$3$:$33$:$03.319$ &$-27$:$41$:$39.037$ &$26.98 \pm 0.21$ &$0.33$ &$0.30$  \\
$95$ &GEMS &$3$:$31$:$49.985$ &$-27$:$51$:$39.640$ &$27.48 \pm 0.32$ &$0.23$ &$0.22$  \\
$96$ &GEMS &$3$:$31$:$48.355$ &$-27$:$58$:$47.325$ &$27.24 \pm 0.21$ &$0.19$ &$0.14$  \\
$98$ &GEMS &$3$:$31$:$26.621$ &$-27$:$44$:$02.180$ &$26.55 \pm 0.14$ &$0.25$ &$0.16$  \\
$99$ &GEMS &$3$:$31$:$40.241$ &$-27$:$45$:$26.825$ &$25.78 \pm 0.07$ &$0.22$ &$0.18$  \\
$100$ &GEMS &$3$:$31$:$54.279$ &$-27$:$58$:$03.475$ &$27.55 \pm 0.35$ &$0.39$ &$0.12$  \\
$101$ &GEMS &$3$:$33$:$07.750$ &$-27$:$38$:$19.356$ &$26.61 \pm 0.15$ &$0.40$ &$0.11$  \\
$105$ &GEMS &$3$:$33$:$12.402$ &$-27$:$45$:$24.309$ &$26.28 \pm 0.11$ &$0.08$ &$0.10$  \\
$106$ &GEMS &$3$:$32$:$21.285$ &$-27$:$36$:$21.341$ &$25.62 \pm 0.06$ &$0.46$ &$0.17$  \\
$109$ &GEMS &$3$:$32$:$08.599$ &$-27$:$57$:$11.897$ &$28.98 \pm 1.25$ &$0.03$ &$0.04$  \\
$112$ &GEMS &$3$:$32$:$44.073$ &$-27$:$37$:$17.813$ &$26.41 \pm 0.12$ &$0.27$ &$0.12$  \\
$113$ &GEMS &$3$:$31$:$35.944$ &$-27$:$50$:$52.915$ &$26.57 \pm 0.14$ &$0.24$ &$0.23$  \\
$114$ &GEMS &$3$:$32$:$10.528$ &$-27$:$59$:$17.638$ &$27.32 \pm 0.28$ &$0.37$ &$0.07$  \\
$115$ &GEMS &$3$:$32$:$24.663$ &$-28$:$01$:$53.101$ &$28.70 \pm 1.00$ &$0.17$ &$0.12$  \\
$121$ &GEMS &$3$:$33$:$23.709$ &$-27$:$44$:$09.139$ &$26.97 \pm 0.21$ &$0.31$ &$0.09$  \\
$122$ &GEMS &$3$:$32$:$20.465$ &$-27$:$35$:$01.627$ &$26.48 \pm 0.13$ &$0.37$ &$0.10$  \\
$124$ &GEMS &$3$:$31$:$42.925$ &$-28$:$03$:$07.798$ &$26.01 \pm 0.08$ &$0.42$ &$0.16$  \\
$126$ &GEMS &$3$:$31$:$44.373$ &$-27$:$50$:$57.703$ &$26.29 \pm 0.11$ &$0.52$ &$0.12$  \\
$127$ &GEMS &$3$:$33$:$02.820$ &$-27$:$57$:$17.505$ &$26.94 \pm 0.20$ &$0.45$ &$0.10$  \\
$128$ &GEMS &$3$:$32$:$30.616$ &$-28$:$00$:$47.901$ &$......$ &$0.25$ &$...$  \\
$129$ &GEMS &$3$:$33$:$15.707$ &$-28$:$02$:$19.075$ &$28.66 \pm 0.95$ &$0.05$ &$0.24$  \\
$132$ &GEMS &$3$:$32$:$10.122$ &$-27$:$53$:$03.530$ &$......$ &$...$ &$...$  \\
$133$ &GEMS &$3$:$31$:$42.953$ &$-27$:$45$:$06.566$ &$27.12 \pm 0.23$ &$0.36$ &$0.12$  \\
$134$ &GEMS &$3$:$31$:$31.591$ &$-27$:$47$:$07.650$ &$27.43 \pm 0.31$ &$...$ &$...$  \\
$137$ &GEMS &$3$:$32$:$59.993$ &$-27$:$50$:$29.076$ &$28.81 \pm 1.09$ &$0.64$ &$0.05$  \\
$138$ &GEMS &$3$:$32$:$54.646$ &$-27$:$38$:$54.076$ &$27.40 \pm 0.30$ &$0.14$ &$0.18$  \\
$140$ &GEMS &$3$:$33$:$24.452$ &$-27$:$44$:$34.000$ &$27.01 \pm 0.22$ &$0.63$ &$0.28$  \\
$141$ &GEMS &$3$:$33$:$02.074$ &$-27$:$51$:$07.660$ &$28.66 \pm 0.98$ &$...$ &$...$  \\
$144$ &GEMS &$3$:$32$:$47.581$ &$-27$:$55$:$38.250$ &$27.26 \pm 0.26$ &$...$ &$...$  \\
$145$ &GEMS &$3$:$32$:$21.539$ &$-27$:$36$:$04.378$ &$27.29 \pm 0.27$ &$0.49$ &$0.21$  \\
$147$ &GEMS &$3$:$31$:$44.967$ &$-27$:$58$:$28.444$ &$28.45 \pm 0.81$ &$0.24$ &$0.07$  \\
$148$ &GEMS &$3$:$32$:$42.738$ &$-27$:$39$:$39.981$ &$27.53 \pm 0.34$ &$0.34$ &$0.07$  \\
$149$ &GEMS &$3$:$33$:$07.027$ &$-27$:$37$:$53.512$ &$26.37 \pm 0.12$ &$0.48$ &$0.16$  \\
$150$ &GEMS &$3$:$33$:$16.894$ &$-27$:$43$:$53.100$ &$......$ &$...$ &$...$  \\
$151$ &GEMS &$3$:$32$:$49.745$ &$-27$:$46$:$43.785$ &$28.30 \pm 0.68$ &$0.34$ &$0.09$  \\
$152$ &GEMS &$3$:$33$:$29.304$ &$-27$:$36$:$41.782$ &$25.93 \pm 0.08$ &$0.52$ &$0.29$  \\
$154$ &GEMS &$3$:$31$:$53.538$ &$-27$:$47$:$00.030$ &$28.45 \pm 0.78$ &$...$ &$...$  \\
$155$ &GEMS &$3$:$32$:$22.267$ &$-27$:$34$:$58.630$ &$28.36 \pm 0.74$ &$...$ &$...$  \\
$156$ &GEMS &$3$:$33$:$00.604$ &$-28$:$00$:$06.537$ &$26.88 \pm 0.18$ &$0.25$ &$0.11$  \\
$157$ &GEMS &$3$:$33$:$28.389$ &$-27$:$45$:$09.634$ &$26.84 \pm 0.19$ &$0.35$ &$0.22$  \\
$159$ &GEMS &$3$:$31$:$29.732$ &$-27$:$50$:$10.500$ &$27.32 \pm 0.27$ &$...$ &$...$  \\
$160$ &GEMS &$3$:$32$:$02.796$ &$-27$:$45$:$28.796$ &$27.65 \pm 0.39$ &$0.27$ &$0.06$  \\
$161$ &GEMS &$3$:$33$:$02.969$ &$-27$:$50$:$29.420$ &$27.44 \pm 0.31$ &$...$ &$...$  \\
$162$ &GEMS &$3$:$33$:$15.184$ &$-27$:$54$:$01.642$ &$25.74 \pm 0.07$ &$0.44$ &$0.30$  \\
\enddata

\tablenotetext{a}{Index from table 2 of Gronwall et al. 2007}
\tablenotetext{b}{Position of ACS centroid (set to ground-based position when there are no SExtractor detections)}
\tablenotetext{c}{Distance between ACS and ground-based centroids}
\tablenotetext{d}{Half-light radius computed by {\tt PHOT} (not reported for LAEs without SExtractor detections)}

\end{deluxetable}

\begin{deluxetable}{lcccccccccc}
\tablecaption{LAE Component Photometric Properties\label{tab:GEMSsex}}
\rotate
\tablewidth{0pt}
\tablehead{
\colhead{Number\tablenotemark{a}}
&\colhead{Component\tablenotemark{b}}
&\colhead{Survey}
&\colhead{$\alpha$}
&\colhead{$\delta$}
&\colhead{$V^{\rm SE}$}
&\colhead{$d_c$\tablenotemark{c}}
&\colhead{$b/a$\tablenotemark{d}}
&\colhead{$\theta$\tablenotemark{e}}
&\colhead{$r_e^{\rm SE}$~\tablenotemark{f}}
&\colhead{SFR(UV)}
\\
&
&
&
&
&
&\colhead{(\arcsec)}
&\colhead{(AB mags)}
&\colhead{($^\circ$)}
&\colhead{(\arcsec)}
&\colhead{(M$_{\sun}$/yr)}
}
\startdata
$25$ &$1$ &HUDF &$3$:$32$:$40.785$ &$-27$:$46$:$06.037$ &$25.04 \pm 0.00$ &$0.14$ &$0.44$ &$-80.30$ &$0.19$ &$4.27$  \\
$56$ &$1$ &HUDF &$3$:$32$:$34.328$ &$-27$:$47$:$59.545$ &$26.37 \pm 0.01$ &$0.35$ &$0.41$ &$-4.20$ &$0.17$ &$1.26$  \\
$125$ &$1$ &HUDF &$3$:$32$:$39.013$ &$-27$:$46$:$22.311$ &$26.56 \pm 0.02$ &$0.44$ &$0.55$ &$-1.20$ &$0.21$ &$1.06$  \\
$4$ &$1$ &GOODS &$3$:$32$:$18.814$ &$-27$:$42$:$48.194$ &$25.27 \pm 0.02$ &$0.02$ &$0.72$ &$48.70$ &$0.11$ &$3.45$  \\
$ $ &$2$ &GOODS &$3$:$32$:$18.786$ &$-27$:$42$:$48.226$ &$27.48 \pm 0.11$ &$0.38$ &$0.76$ &$84.50$ &$0.09$ &$0.45$  \\
$ $ &$3$ &GOODS &$3$:$32$:$18.795$ &$-27$:$42$:$47.868$ &$27.56 \pm 0.13$ &$0.40$ &$0.46$ &$18.60$ &$0.12$ &$0.42$  \\
$ $ &$4$ &GOODS &$3$:$32$:$18.841$ &$-27$:$42$:$47.809$ &$27.62 \pm 0.12$ &$0.50$ &$0.44$ &$-77.50$ &$0.12$ &$0.40$  \\
$ $ &$5$ &GOODS &$3$:$32$:$18.835$ &$-27$:$42$:$47.646$ &$27.96 \pm 0.12$ &$0.59$ &$0.19$ &$-78.60$ &$0.11$ &$0.29$  \\
$6$ &$1$ &GOODS &$3$:$32$:$52.690$ &$-27$:$48$:$09.284$ &$25.44 \pm 0.02$ &$0.19$ &$0.83$ &$-42.10$ &$0.09$ &$2.97$  \\
$11$ &$1$ &GOODS &$3$:$32$:$26.922$ &$-27$:$41$:$28.046$ &$25.59 \pm 0.03$ &$0.54$ &$0.72$ &$53.80$ &$0.17$ &$2.59$  \\
$ $ &$2$ &GOODS &$3$:$32$:$26.969$ &$-27$:$41$:$27.718$ &$26.35 \pm 0.04$ &$0.38$ &$0.90$ &$-65.90$ &$0.10$ &$1.28$  \\
$25$ &$1$ &GOODS &$3$:$32$:$40.785$ &$-27$:$46$:$06.053$ &$25.14 \pm 0.02$ &$0.12$ &$0.42$ &$-74.60$ &$0.17$ &$3.91$  \\
$ $ &$2$ &GOODS &$3$:$32$:$40.784$ &$-27$:$46$:$05.450$ &$27.63 \pm 0.13$ &$0.52$ &$0.74$ &$49.90$ &$0.12$ &$0.40$  \\
$35$ &$1$ &GOODS &$3$:$32$:$45.604$ &$-27$:$52$:$10.914$ &$27.07 \pm 0.06$ &$0.09$ &$0.88$ &$87.50$ &$0.08$ &$0.66$  \\
$41$ &$1$ &GOODS &$3$:$32$:$56.672$ &$-27$:$49$:$49.199$ &$26.83 \pm 0.08$ &$0.11$ &$0.52$ &$-62.10$ &$0.08$ &$0.82$  \\
$44$ &$1$ &GOODS &$3$:$32$:$15.790$ &$-27$:$44$:$09.919$ &$26.51 \pm 0.05$ &$0.24$ &$0.95$ &$-25.50$ &$0.13$ &$1.11$  \\
$ $ &$2$ &GOODS &$3$:$32$:$15.809$ &$-27$:$44$:$10.074$ &$26.59 \pm 0.07$ &$0.16$ &$0.71$ &$28.60$ &$0.19$ &$1.03$  \\
$55$ &$1$ &GOODS &$3$:$32$:$59.972$ &$-27$:$50$:$26.366$ &$26.88 \pm 0.12$ &$0.37$ &$0.74$ &$-18.60$ &$0.09$ &$0.79$  \\
$ $ &$2$ &GOODS &$3$:$32$:$59.985$ &$-27$:$50$:$26.190$ &$27.63 \pm 0.19$ &$0.36$ &$0.47$ &$-49.80$ &$0.09$ &$0.39$  \\
$56$ &$1$ &GOODS &$3$:$32$:$34.331$ &$-27$:$47$:$59.554$ &$26.53 \pm 0.08$ &$0.37$ &$0.29$ &$-16.10$ &$0.17$ &$1.09$  \\
$59$ &$1$ &GOODS &$3$:$32$:$33.254$ &$-27$:$51$:$27.590$ &$25.94 \pm 0.05$ &$0.31$ &$0.46$ &$-5.20$ &$0.16$ &$1.86$  \\
$66$ &$1$ &GOODS &$3$:$32$:$48.528$ &$-27$:$56$:$05.376$ &$27.53 \pm 0.15$ &$0.18$ &$0.79$ &$17.90$ &$0.07$ &$0.43$  \\
$85$ &$1$ &GOODS &$3$:$32$:$59.824$ &$-27$:$53$:$05.768$ &$26.98 \pm 0.07$ &$0.25$ &$0.91$ &$3.90$ &$0.07$ &$0.72$  \\
$90$ &$1$ &GOODS &$3$:$32$:$14.574$ &$-27$:$45$:$52.417$ &$27.47 \pm 0.08$ &$0.10$ &$0.73$ &$38.20$ &$0.06$ &$0.46$  \\
$94$ &$1$ &GOODS &$3$:$32$:$09.336$ &$-27$:$43$:$54.193$ &$26.72 \pm 0.05$ &$0.46$ &$0.89$ &$38.20$ &$0.09$ &$0.91$  \\
$97$ &$1$ &GOODS &$3$:$32$:$12.584$ &$-27$:$48$:$05.665$ &$27.91 \pm 0.13$ &$0.60$ &$0.77$ &$-76.80$ &$0.08$ &$0.30$  \\
$102$ &$1$ &GOODS &$3$:$32$:$57.356$ &$-27$:$51$:$42.357$ &$27.85 \pm 0.15$ &$0.26$ &$0.68$ &$73.80$ &$0.08$ &$0.32$  \\
$107$ &$1$ &GOODS &$3$:$32$:$10.158$ &$-27$:$44$:$28.336$ &$27.92 \pm 0.14$ &$0.32$ &$0.44$ &$-9.20$ &$0.08$ &$0.30$  \\
$125$ &$1$ &GOODS &$3$:$32$:$39.023$ &$-27$:$46$:$22.290$ &$27.08 \pm 0.09$ &$0.56$ &$0.73$ &$52.60$ &$0.10$ &$0.65$  \\
$136$ &$1$ &GOODS &$3$:$32$:$24.329$ &$-27$:$41$:$51.886$ &$27.90 \pm 0.11$ &$0.26$ &$0.74$ &$73.40$ &$0.07$ &$0.31$  \\
$141$ &$1$ &GOODS &$3$:$33$:$02.097$ &$-27$:$51$:$07.663$ &$27.73 \pm 0.15$ &$0.26$ &$0.45$ &$79.70$ &$0.06$ &$0.36$  \\
$143$ &$1$ &GOODS &$3$:$32$:$54.652$ &$-27$:$51$:$49.228$ &$28.00 \pm 0.13$ &$0.09$ &$0.72$ &$27.10$ &$0.07$ &$0.28$  \\
$146$ &$1$ &GOODS &$3$:$32$:$44.168$ &$-27$:$50$:$56.842$ &$28.01 \pm 0.13$ &$0.21$ &$0.83$ &$66.60$ &$0.07$ &$0.28$  \\
$160$ &$1$ &GOODS &$3$:$32$:$02.802$ &$-27$:$45$:$28.708$ &$27.74 \pm 0.11$ &$0.15$ &$0.71$ &$-34.40$ &$0.07$ &$0.36$  \\
$5$ &$1$ &GEMS &$3$:$32$:$47.516$ &$-27$:$58$:$07.706$ &$25.01 \pm 0.02$ &$0.11$ &$0.43$ &$77.60$ &$0.14$ &$4.39$  \\
$8$ &$1$ &GEMS &$3$:$31$:$54.884$ &$-27$:$51$:$21.124$ &$25.45 \pm 0.02$ &$0.12$ &$0.77$ &$-53.80$ &$0.08$ &$2.93$  \\
$9$ &$1$ &GEMS &$3$:$31$:$40.157$ &$-28$:$03$:$07.406$ &$25.24 \pm 0.01$ &$0.10$ &$0.82$ &$85.30$ &$0.07$ &$3.57$  \\
$10$ &$1$ &GEMS &$3$:$33$:$22.460$ &$-27$:$46$:$36.923$ &$25.87 \pm 0.03$ &$0.06$ &$0.74$ &$81.90$ &$0.08$ &$1.99$  \\
$ $ &$2$ &GEMS &$3$:$33$:$22.424$ &$-27$:$46$:$36.777$ &$25.96 \pm 0.07$ &$0.45$ &$0.41$ &$83.60$ &$0.22$ &$1.83$  \\
$12$ &$1$ &GEMS &$3$:$32$:$33.838$ &$-27$:$36$:$35.223$ &$26.05 \pm 0.05$ &$0.03$ &$0.66$ &$-68.60$ &$0.11$ &$1.69$  \\
$ $ &$2$ &GEMS &$3$:$32$:$33.863$ &$-27$:$36$:$34.885$ &$26.85 \pm 0.09$ &$0.50$ &$0.75$ &$-81.80$ &$0.16$ &$0.81$  \\
$13$ &$1$ &GEMS &$3$:$33$:$07.252$ &$-27$:$47$:$47.176$ &$25.65 \pm 0.04$ &$0.04$ &$0.39$ &$-45.80$ &$0.11$ &$2.45$  \\
$15$ &$1$ &GEMS &$3$:$33$:$18.925$ &$-27$:$38$:$28.474$ &$26.49 \pm 0.07$ &$0.27$ &$0.74$ &$-49.60$ &$0.10$ &$1.12$  \\
$ $ &$2$ &GEMS &$3$:$33$:$18.894$ &$-27$:$38$:$28.456$ &$27.37 \pm 0.13$ &$0.58$ &$0.53$ &$68.30$ &$0.13$ &$0.50$  \\
$17$ &$1$ &GEMS &$3$:$32$:$49.148$ &$-27$:$34$:$39.971$ &$25.67 \pm 0.04$ &$0.38$ &$0.74$ &$-85.80$ &$0.12$ &$2.40$  \\
$18$ &$1$ &GEMS &$3$:$32$:$46.754$ &$-27$:$39$:$59.914$ &$26.77 \pm 0.06$ &$0.18$ &$0.92$ &$-78.20$ &$0.08$ &$0.87$  \\
$19$ &$1$ &GEMS &$3$:$31$:$34.738$ &$-27$:$56$:$21.821$ &$25.30 \pm 0.02$ &$0.41$ &$0.93$ &$18.40$ &$0.08$ &$3.38$  \\
$20$ &$1$ &GEMS &$3$:$33$:$11.883$ &$-28$:$00$:$12.572$ &$26.47 \pm 0.09$ &$0.24$ &$0.82$ &$71.20$ &$0.15$ &$1.15$  \\
$ $ &$2$ &GEMS &$3$:$33$:$11.873$ &$-28$:$00$:$12.142$ &$26.67 \pm 0.07$ &$0.58$ &$0.84$ &$86.50$ &$0.09$ &$0.95$  \\
$22$ &$1$ &GEMS &$3$:$31$:$51.636$ &$-27$:$58$:$32.642$ &$26.41 \pm 0.04$ &$0.06$ &$0.70$ &$41.80$ &$0.08$ &$1.21$  \\
$24$ &$1$ &GEMS &$3$:$31$:$53.213$ &$-27$:$57$:$08.160$ &$26.29 \pm 0.04$ &$0.07$ &$0.85$ &$50.70$ &$0.07$ &$1.36$  \\
$26$ &$1$ &GEMS &$3$:$31$:$51.561$ &$-27$:$46$:$47.016$ &$26.53 \pm 0.08$ &$0.38$ &$0.66$ &$-12.50$ &$0.11$ &$1.08$  \\
$29$ &$1$ &GEMS &$3$:$31$:$47.796$ &$-27$:$45$:$03.320$ &$26.75 \pm 0.07$ &$0.34$ &$0.89$ &$47.90$ &$0.09$ &$0.88$  \\
$ $ &$2$ &GEMS &$3$:$31$:$47.821$ &$-27$:$45$:$03.434$ &$28.99 \pm 0.26$ &$0.39$ &$0.52$ &$-17.00$ &$0.05$ &$0.11$  \\
$32$ &$1$ &GEMS &$3$:$33$:$19.793$ &$-27$:$38$:$20.505$ &$27.17 \pm 0.08$ &$0.19$ &$0.77$ &$58.70$ &$0.08$ &$0.60$  \\
$36$ &$1$ &GEMS &$3$:$32$:$18.926$ &$-27$:$38$:$40.195$ &$27.11 \pm 0.12$ &$0.21$ &$0.84$ &$-63.20$ &$0.09$ &$0.64$  \\
$38$ &$1$ &GEMS &$3$:$31$:$50.369$ &$-27$:$59$:$10.119$ &$26.55 \pm 0.05$ &$0.03$ &$0.86$ &$-59.00$ &$0.08$ &$1.07$  \\
$39$ &$1$ &GEMS &$3$:$31$:$30.524$ &$-27$:$47$:$29.647$ &$25.45 \pm 0.03$ &$0.35$ &$0.45$ &$-54.10$ &$0.18$ &$2.93$  \\
$40$ &$1$ &GEMS &$3$:$31$:$31.477$ &$-27$:$34$:$47.266$ &$27.66 \pm 0.13$ &$0.06$ &$0.63$ &$-43.30$ &$0.07$ &$0.39$  \\
$ $ &$2$ &GEMS &$3$:$31$:$31.454$ &$-27$:$34$:$47.849$ &$28.94 \pm 0.25$ &$0.59$ &$0.40$ &$41.80$ &$0.05$ &$0.12$  \\
$41$ &$1$ &GEMS &$3$:$32$:$56.672$ &$-27$:$49$:$49.241$ &$26.88 \pm 0.08$ &$0.15$ &$0.72$ &$-83.10$ &$0.09$ &$0.79$  \\
$43$ &$1$ &GEMS &$3$:$33$:$07.315$ &$-27$:$54$:$38.983$ &$25.93 \pm 0.04$ &$0.19$ &$0.92$ &$-60.70$ &$0.08$ &$1.88$  \\
$49$ &$1$ &GEMS &$3$:$31$:$42.359$ &$-27$:$58$:$07.857$ &$25.76 \pm 0.05$ &$0.44$ &$0.57$ &$-30.60$ &$0.15$ &$2.20$  \\
$50$ &$1$ &GEMS &$3$:$31$:$52.830$ &$-27$:$45$:$18.650$ &$27.09 \pm 0.07$ &$0.33$ &$0.51$ &$-22.20$ &$0.08$ &$0.65$  \\
$52$ &$1$ &GEMS &$3$:$33$:$21.363$ &$-27$:$38$:$36.337$ &$27.05 \pm 0.07$ &$0.43$ &$0.82$ &$52.00$ &$0.06$ &$0.67$  \\
$53$ &$1$ &GEMS &$3$:$32$:$15.131$ &$-27$:$38$:$53.946$ &$25.87 \pm 0.04$ &$0.46$ &$0.58$ &$47.40$ &$0.16$ &$2.00$  \\
$55$ &$1$ &GEMS &$3$:$32$:$59.974$ &$-27$:$50$:$26.401$ &$26.71 \pm 0.11$ &$0.31$ &$0.54$ &$-77.30$ &$0.14$ &$0.92$  \\
$ $ &$2$ &GEMS &$3$:$32$:$59.988$ &$-27$:$50$:$26.204$ &$27.61 \pm 0.13$ &$0.33$ &$0.69$ &$2.00$ &$0.08$ &$0.40$  \\
$58$ &$1$ &GEMS &$3$:$33$:$06.943$ &$-27$:$42$:$27.852$ &$27.08 \pm 0.09$ &$0.16$ &$0.71$ &$81.30$ &$0.08$ &$0.66$  \\
$61$ &$1$ &GEMS &$3$:$33$:$09.431$ &$-27$:$45$:$50.096$ &$26.40 \pm 0.06$ &$0.47$ &$0.51$ &$75.50$ &$0.14$ &$1.22$  \\
$ $ &$2$ &GEMS &$3$:$33$:$09.398$ &$-27$:$45$:$50.025$ &$28.04 \pm 0.12$ &$0.27$ &$0.84$ &$-77.60$ &$0.08$ &$0.27$  \\
$ $ &$3$ &GEMS &$3$:$33$:$09.402$ &$-27$:$45$:$50.235$ &$28.59 \pm 0.14$ &$0.06$ &$0.60$ &$-2.50$ &$0.05$ &$0.16$  \\
$63$ &$1$ &GEMS &$3$:$32$:$51.917$ &$-27$:$42$:$12.246$ &$25.81 \pm 0.03$ &$0.38$ &$0.64$ &$67.60$ &$0.09$ &$2.10$  \\
$64$ &$1$ &GEMS &$3$:$31$:$59.831$ &$-27$:$49$:$46.424$ &$26.41 \pm 0.06$ &$0.16$ &$0.83$ &$53.40$ &$0.09$ &$1.21$  \\
$66$ &$1$ &GEMS &$3$:$32$:$48.527$ &$-27$:$56$:$05.357$ &$27.92 \pm 0.15$ &$0.18$ &$0.73$ &$-40.00$ &$0.06$ &$0.30$  \\
$67$ &$1$ &GEMS &$3$:$32$:$51.770$ &$-27$:$37$:$33.552$ &$26.36 \pm 0.04$ &$0.02$ &$0.91$ &$0.80$ &$0.08$ &$1.27$  \\
$68$ &$1$ &GEMS &$3$:$32$:$58.130$ &$-27$:$48$:$04.885$ &$27.20 \pm 0.12$ &$0.33$ &$0.73$ &$-42.00$ &$0.10$ &$0.59$  \\
$ $ &$2$ &GEMS &$3$:$32$:$58.159$ &$-27$:$48$:$04.861$ &$27.95 \pm 0.15$ &$0.20$ &$0.63$ &$45.20$ &$0.07$ &$0.29$  \\
$69$ &$1$ &GEMS &$3$:$33$:$25.357$ &$-28$:$02$:$46.531$ &$26.11 \pm 0.05$ &$0.13$ &$0.67$ &$-23.80$ &$0.13$ &$1.60$  \\
$72$ &$1$ &GEMS &$3$:$33$:$07.179$ &$-27$:$48$:$51.544$ &$27.41 \pm 0.11$ &$0.20$ &$0.73$ &$-55.30$ &$0.07$ &$0.48$  \\
$73$ &$1$ &GEMS &$3$:$32$:$57.404$ &$-27$:$55$:$19.073$ &$26.59 \pm 0.06$ &$0.44$ &$0.89$ &$34.60$ &$0.08$ &$1.02$  \\
$74$ &$1$ &GEMS &$3$:$33$:$18.589$ &$-27$:$45$:$42.588$ &$26.78 \pm 0.05$ &$0.26$ &$0.83$ &$80.30$ &$0.08$ &$0.86$  \\
$ $ &$2$ &GEMS &$3$:$33$:$18.575$ &$-27$:$45$:$42.755$ &$29.46 \pm 0.42$ &$0.39$ &$0.63$ &$47.50$ &$0.05$ &$0.07$  \\
$ $ &$3$ &GEMS &$3$:$33$:$18.584$ &$-27$:$45$:$42.826$ &$29.87 \pm 0.46$ &$0.30$ &$0.52$ &$-62.50$ &$0.03$ &$0.05$  \\
$75$ &$1$ &GEMS &$3$:$32$:$59.268$ &$-27$:$41$:$14.756$ &$27.12 \pm 0.09$ &$0.27$ &$0.56$ &$37.20$ &$0.08$ &$0.63$  \\
$78$ &$1$ &GEMS &$3$:$33$:$20.840$ &$-27$:$51$:$45.883$ &$27.54 \pm 0.23$ &$0.30$ &$0.97$ &$-42.40$ &$0.11$ &$0.43$  \\
$79$ &$1$ &GEMS &$3$:$31$:$58.025$ &$-27$:$47$:$30.314$ &$27.14 \pm 0.07$ &$0.46$ &$0.70$ &$-45.20$ &$0.08$ &$0.62$  \\
$82$ &$1$ &GEMS &$3$:$31$:$47.776$ &$-27$:$42$:$16.325$ &$26.39 \pm 0.05$ &$0.36$ &$0.78$ &$72.20$ &$0.09$ &$1.23$  \\
$83$ &$1$ &GEMS &$3$:$31$:$38.669$ &$-27$:$45$:$43.559$ &$26.80 \pm 0.06$ &$0.01$ &$0.81$ &$72.90$ &$0.10$ &$0.84$  \\
$86$ &$1$ &GEMS &$3$:$33$:$28.170$ &$-28$:$03$:$20.782$ &$27.47 \pm 0.09$ &$0.33$ &$0.95$ &$-34.80$ &$0.07$ &$0.46$  \\
$87$ &$1$ &GEMS &$3$:$33$:$05.027$ &$-27$:$43$:$37.306$ &$26.31 \pm 0.06$ &$0.20$ &$0.56$ &$-39.80$ &$0.11$ &$1.32$  \\
$89$ &$1$ &GEMS &$3$:$33$:$12.017$ &$-27$:$58$:$39.929$ &$27.11 \pm 0.12$ &$0.49$ &$0.38$ &$87.50$ &$0.11$ &$0.64$  \\
$91$ &$1$ &GEMS &$3$:$31$:$58.803$ &$-27$:$49$:$28.767$ &$26.85 \pm 0.07$ &$0.46$ &$0.67$ &$-68.80$ &$0.09$ &$0.81$  \\
$92$ &$1$ &GEMS &$3$:$33$:$03.327$ &$-27$:$41$:$38.828$ &$28.31 \pm 0.15$ &$0.09$ &$0.58$ &$82.10$ &$0.05$ &$0.21$  \\
$95$ &$1$ &GEMS &$3$:$31$:$49.991$ &$-27$:$51$:$39.476$ &$28.02 \pm 0.18$ &$0.22$ &$0.59$ &$54.00$ &$0.08$ &$0.28$  \\
$96$ &$1$ &GEMS &$3$:$31$:$48.358$ &$-27$:$58$:$47.429$ &$27.51 \pm 0.09$ &$0.21$ &$0.70$ &$-69.10$ &$0.10$ &$0.44$  \\
$ $ &$2$ &GEMS &$3$:$31$:$48.352$ &$-27$:$58$:$47.229$ &$27.52 \pm 0.13$ &$0.11$ &$0.79$ &$46.50$ &$0.12$ &$0.44$  \\
$98$ &$1$ &GEMS &$3$:$31$:$26.620$ &$-27$:$44$:$02.186$ &$26.85 \pm 0.10$ &$0.26$ &$0.59$ &$25.20$ &$0.13$ &$0.81$  \\
$99$ &$1$ &GEMS &$3$:$31$:$40.247$ &$-27$:$45$:$26.715$ &$26.08 \pm 0.04$ &$0.38$ &$0.69$ &$44.00$ &$0.10$ &$1.64$  \\
$ $ &$2$ &GEMS &$3$:$31$:$40.214$ &$-27$:$45$:$27.357$ &$28.09 \pm 0.20$ &$0.41$ &$0.55$ &$-8.80$ &$0.09$ &$0.26$  \\
$100$ &$1$ &GEMS &$3$:$31$:$54.279$ &$-27$:$58$:$03.475$ &$28.14 \pm 0.19$ &$0.34$ &$0.43$ &$-54.40$ &$0.07$ &$0.25$  \\
$101$ &$1$ &GEMS &$3$:$33$:$07.750$ &$-27$:$38$:$19.356$ &$26.82 \pm 0.09$ &$0.37$ &$0.57$ &$19.70$ &$0.09$ &$0.83$  \\
$105$ &$1$ &GEMS &$3$:$33$:$12.402$ &$-27$:$45$:$24.309$ &$26.31 \pm 0.06$ &$0.02$ &$0.82$ &$-71.00$ &$0.10$ &$1.33$  \\
$106$ &$1$ &GEMS &$3$:$32$:$21.284$ &$-27$:$36$:$21.415$ &$25.86 \pm 0.05$ &$0.49$ &$0.71$ &$-51.60$ &$0.12$ &$2.01$  \\
$ $ &$2$ &GEMS &$3$:$32$:$21.291$ &$-27$:$36$:$21.046$ &$27.36 \pm 0.15$ &$0.39$ &$0.54$ &$79.00$ &$0.14$ &$0.51$  \\
$112$ &$1$ &GEMS &$3$:$32$:$44.073$ &$-27$:$37$:$17.813$ &$26.44 \pm 0.06$ &$0.30$ &$0.48$ &$-14.50$ &$0.12$ &$1.18$  \\
$113$ &$1$ &GEMS &$3$:$31$:$35.944$ &$-27$:$50$:$52.915$ &$27.36 \pm 0.12$ &$0.20$ &$0.73$ &$-80.50$ &$0.09$ &$0.51$  \\
$114$ &$1$ &GEMS &$3$:$32$:$10.528$ &$-27$:$59$:$17.638$ &$27.17 \pm 0.09$ &$0.36$ &$0.61$ &$10.90$ &$0.08$ &$0.60$  \\
$121$ &$1$ &GEMS &$3$:$33$:$23.709$ &$-27$:$44$:$09.139$ &$27.11 \pm 0.10$ &$0.32$ &$0.56$ &$-81.30$ &$0.08$ &$0.64$  \\
$122$ &$1$ &GEMS &$3$:$32$:$20.465$ &$-27$:$35$:$01.627$ &$26.54 \pm 0.06$ &$0.34$ &$0.61$ &$-75.30$ &$0.09$ &$1.07$  \\
$124$ &$1$ &GEMS &$3$:$31$:$42.923$ &$-28$:$03$:$07.855$ &$26.38 \pm 0.06$ &$0.45$ &$0.51$ &$59.00$ &$0.10$ &$1.25$  \\
$ $ &$2$ &GEMS &$3$:$31$:$42.949$ &$-28$:$03$:$07.277$ &$28.90 \pm 0.31$ &$0.26$ &$0.48$ &$-13.30$ &$0.06$ &$0.12$  \\
$ $ &$3$ &GEMS &$3$:$31$:$42.943$ &$-28$:$03$:$07.203$ &$29.57 \pm 0.37$ &$0.26$ &$0.53$ &$-42.90$ &$0.07$ &$0.07$  \\
$126$ &$1$ &GEMS &$3$:$31$:$44.373$ &$-27$:$50$:$57.703$ &$26.59 \pm 0.06$ &$0.48$ &$0.83$ &$-59.10$ &$0.08$ &$1.03$  \\
$127$ &$1$ &GEMS &$3$:$33$:$02.820$ &$-27$:$57$:$17.505$ &$27.17 \pm 0.10$ &$0.39$ &$0.75$ &$12.10$ &$0.08$ &$0.60$  \\
$133$ &$1$ &GEMS &$3$:$31$:$42.953$ &$-27$:$45$:$06.566$ &$27.07 \pm 0.12$ &$0.31$ &$0.50$ &$-57.10$ &$0.13$ &$0.66$  \\
$138$ &$1$ &GEMS &$3$:$32$:$54.646$ &$-27$:$38$:$54.076$ &$28.22 \pm 0.23$ &$0.16$ &$0.40$ &$53.20$ &$0.07$ &$0.23$  \\
$140$ &$1$ &GEMS &$3$:$33$:$24.452$ &$-27$:$44$:$34.000$ &$28.24 \pm 0.20$ &$0.60$ &$0.54$ &$72.30$ &$0.07$ &$0.23$  \\
$145$ &$1$ &GEMS &$3$:$32$:$21.541$ &$-27$:$36$:$04.470$ &$27.84 \pm 0.18$ &$0.42$ &$0.59$ &$-31.10$ &$0.08$ &$0.32$  \\
$148$ &$1$ &GEMS &$3$:$32$:$42.738$ &$-27$:$39$:$39.981$ &$27.44 \pm 0.08$ &$0.27$ &$0.74$ &$-79.00$ &$0.07$ &$0.47$  \\
$149$ &$1$ &GEMS &$3$:$33$:$07.027$ &$-27$:$37$:$53.512$ &$26.81 \pm 0.09$ &$0.46$ &$0.73$ &$-82.00$ &$0.10$ &$0.84$  \\
$152$ &$1$ &GEMS &$3$:$33$:$29.304$ &$-27$:$36$:$41.782$ &$27.91 \pm 0.20$ &$0.50$ &$0.28$ &$75.20$ &$0.12$ &$0.30$  \\
$156$ &$1$ &GEMS &$3$:$33$:$00.604$ &$-28$:$00$:$06.537$ &$26.99 \pm 0.09$ &$0.22$ &$0.58$ &$66.60$ &$0.10$ &$0.71$  \\
$157$ &$1$ &GEMS &$3$:$33$:$28.389$ &$-27$:$45$:$09.634$ &$27.47 \pm 0.15$ &$0.35$ &$0.83$ &$25.30$ &$0.08$ &$0.46$  \\
$160$ &$1$ &GEMS &$3$:$32$:$02.796$ &$-27$:$45$:$28.796$ &$27.47 \pm 0.11$ &$0.28$ &$0.63$ &$-38.80$ &$0.07$ &$0.46$  \\
$162$ &$1$ &GEMS &$3$:$33$:$15.187$ &$-27$:$54$:$01.865$ &$26.43 \pm 0.07$ &$0.55$ &$0.72$ &$71.70$ &$0.11$ &$1.19$  \\
$ $ &$2$ &GEMS &$3$:$33$:$15.185$ &$-27$:$54$:$01.566$ &$26.84 \pm 0.09$ &$0.35$ &$0.58$ &$77.60$ &$0.18$ &$0.82$  \\
$ $ &$3$ &GEMS &$3$:$33$:$15.176$ &$-27$:$54$:$01.270$ &$27.27 \pm 0.13$ &$0.44$ &$0.58$ &$1.50$ &$0.12$ &$0.55$  \\
\enddata

\tablenotetext{a}{Index from table 2 of Gronwall et al. 2007}
\tablenotetext{b}{Component number}
\tablenotetext{c}{Distance from ground-based Ly$\alpha$ position}
\tablenotetext{d}{Isophotal axis ratio computed by SExtractor}
\tablenotetext{e}{Isophotal position angle computed by SExtractor}
\tablenotetext{f}{Half-light radius computed by SExtractor}

\end{deluxetable}

\end{document}